\documentclass[12pt]{article}
\usepackage[letterpaper,margin=2cm,includehead,nomarginpar,headheight=2cm]{geometry}
\usepackage{amsmath,amssymb,fancyhdr,graphicx,xcolor,float,titling,scalerel}
\usepackage{hyperref,sectsty,lastpage,soul}
\usepackage{diagbox,subcaption,multicol,tabularx,booktabs,lipsum}
\usepackage{tikz}
\usepackage{accents}
\usepackage{multirow}
\usepackage{lineno}
\hypersetup{breaklinks=true, bookmarks=true, pdfauthor={Valentin Vergara Hidd}, colorlinks=true, citecolor=gmu-green, urlcolor=gmu-green, linkcolor=gmu-coral, pdfborder={0 0 0}}
\definecolor{gmu-green}{RGB}{30,98,56}
\definecolor{gmu-gold}{RGB}{226,168,43}
\definecolor{gmu-coral}{RGB}{243,112,33}
\definecolor{wu_color}{HTML}{EE7733}
\definecolor{uic_color}{HTML}{0077BB}
\definecolor{ocs_color}{HTML}{BBBBBB}
\definecolor{wul_color}{HTML}{EE3377}
\definecolor{all_uic_color}{HTML}{AA4489}
\definecolor{all_ocs_color}{HTML}{009988}
\definecolor{sl_wu_color}{HTML}{33BBEE}
\definecolor{alt_green}{HTML}{005f73}
\definecolor{wu_no_wage}{HTML}{CC0000}
\definecolor{wu_wage_increase}{HTML}{228B22}
\definecolor{wu_wage_decrease}{HTML}{8B4513}

\title{\sc Organizations, teams, and job mobility: A social microdynamics approach inspired by a large US organization}
\author{Bryan Adams \thanks{George Mason University. \href{mailto:badams29@gmu.edu}{badams29@gmu.edu}} \and
  Valent\'{i}n Vergara Hidd \thanks{George Mason University. \href{mailto:vvergara@gmu.edu}{vvergara@gmu.edu}} \and
  Daniel Stimpson \thanks{United States Army Acquisition Support Center (USAASC), 9900 Belvoir Road, Fort Belvoir, VA 22060, USA} \and
  Miesha Purcell \thanks{United States Army Acquisition Support Center (USAASC), 9900 Belvoir Road, Fort Belvoir, VA 22060, USA} \and
  Eduardo L\'{o}pez \thanks{George Mason University. \href{mailto:elopez22@gmu.edu}{elopez22@gmu.edu}}
}

\date{\today}

\makeatletter
\DeclareRobustCommand{\cev}[1]{%
  \mathpalette\do@cev{#1}%
}
\newcommand{\do@cev}[2]{%
  \fix@cev{#1}{+}%
  \reflectbox{$\m@th#1\vec{\reflectbox{$\fix@cev{#1}{-}\m@th#1#2\fix@cev{#1}{+}$}}$}%
  \fix@cev{#1}{-}%
}
\newcommand{\fix@cev}[2]{%
  \ifx#1\displaystyle
    \mkern#23mu
  \else
    \ifx#1\textstyle
      \mkern#23mu
    \else
      \ifx#1\scriptstyle
        \mkern#22mu
      \else
        \mkern#22mu
      \fi
    \fi
  \fi
}

\makeatother

\usepackage{lastpage}

\begin{document}
\maketitle

\begin{abstract}
Most of the modeling approaches used to understand organizational worker mobility are highly stylized, using idealizations such as structureless organizations, indistinguishable workers, and a lack of social bonding of the workers. In this article, aided by a decade of precise, temporally resolved data of a large civilian organization of the US Army in which employees can change jobs in a similar way to many private organizations, we introduce a new framework to describe organizations as composites of teams within which individuals perform specific tasks and where social connections develop. By tracking the personnel composition of organizational teams, we find that workers who change jobs are highly influenced by preferring to reunite with past coworkers. In this organization, 34\% of all moves across temporally stable teams (and 32\% of the totality of moves) lead to worker reunions, percentages that have not been reported and are well-above intuitive expectation. To assess the importance of worker reunions in determining job moves, we compare them to labor supply and demand with or without occupational specialization. The comparison shows that the most consistent information about job change is provided by reunions. We find that the greater the time workers spend together or the smaller the team they share both increase their likelihood to reunite, supporting the notion of increased familiarity and trust behind such reunions and the dominant role of social capital in the evolution of large organizations. Our study of this organization supports the idea that to correctly forecast job mobility inside large organizations, their teams structures and the social ties formed in those teams play a key role in shaping internal job change.
\end{abstract}

\newpage

\section{Main}
Organizations such as companies, institutions, or governmental departments, especially if they are large, are permanently concerned with the best management of their personnel, including their mobility and progression through the organization~\cite{bidwell2020no}. Over the decades, many disciplines have studied the problem of employee organizational mobility with a variety of conceptual frameworks, techniques, and levels of resolution~\cite{manpower,personnel_economics,labor_flows}. While a large portion of the research has been dedicated to the causes and consequences of individual job transition, the system-level question of the collection of detailed patterns of job mobility within an organization has mostly been studied in the manpower analysis and modeling literature~\cite{smith1988manpower,robbert2024modeling}. Manpower models have provided an important tool with which to plan personnel recruitment, promotion, and retention strategies. However, the models also rest on simplifications that selectively ignore many of the specifics characterizing real organizations and their employees, such as the way personnel is structured within the organization (e.g. teams), individuality of knowledge, skills, and abilities, and the social interactions that develop among coworkers and teammates, among others.

It should be noted that manpower analysis originated with the benefit of the organization in mind (i.e. organization-centric). However, as the models began to be adapted to the perspective of the employees (individual-centric), a number of important realizations arose such as the impact on individual outcomes of what is called the organization's \textit{mobility process} (i.e. the rules by which people can change jobs within the organization)~\cite{bidwell2020no}. Among the possible mobility processes, so-called \textit{vacancy systems} are the most precisely modeled~\cite{white_1970,barth_stat_manpower,stewman,chase1991vacancy}. In vacancy systems, employees typically can take a new position when a vacancy is made available by the organization, most commonly because the position has been left vacant by another employee. Numerous studies have explored this system, uncovering important patterns affecting internal labor markets such as chains of vacancies that generate system-level job progression~\cite{white_1970,chase1991vacancy}, pinch-points limiting employees' organizational advancement~\cite{stewman1983careers}, and the fact that job progression can operate as a tournament among employees~\cite{rosenbaum_tournaments}. A great advantage of \textit{vacancy systems models} (VSMs), as with all manpower analysis, is their micro-macro capability, modeling the full scope of the organizational workforce starting from the individual level. Yet, despite their successes, these models inherit from manpower models the same simplifying assumptions mentioned above.

A particularly glaring shortcoming of VSMs and, in general, individual-centric system-wide organizational mobility models, is the lack of consideration for social interactions. While the latter have become a major topic of interest in the labor literature both for open and internal labor markets~\cite{jackson2008social,podolny1997resources}, individual-centric system-wide organizational mobility models do not incorporate them. To be fair, inclusion of social effects on these models requires understanding of the social structure of an organization in order to capture relevant social ties. This, in turn, requires precise information about an organization's internal structure, information that is not usually available. Furthermore, there may be a question of relevance as well: how meaningful can the inclusion of social interaction into VSMs or other precise mobility models be?   

To make progress in this question, we present empirical evidence for the existence of a strong effect of organizational social interactions on organizational job mobility. Studying the case of a large organization, we find an unexpected abundance of job changes that reunite previous coworkers. This effect is detected using a novel method to construct the temporal team structure of the organization on the basis of monthly personnel data covering a span of nine years (beginning of 2012 to end of 2020). When individuals work together in a team at any point in time, they are considered coworkers and hence to have interacted socially. We show here that the abundance of reuniting moves in the organization studied (34\% of moves between stable teams over time) far exceeds what is expected based on other features of job mobility such as labor supply and demand or occupational specialization. Furthermore, we find evidence that the likelihood of reuniting moves is affected by the context of interaction in which reuniting coworkers previously spent time together. Namely, coworkers that spent more time together and in smaller teams are more likely to be reunited. As we argue later, we believe that one mechanism by which social interactions lead to reuniting moves are internal job referrals~\cite{referrals-review}. To assess how important reuniting moves are in shaping job mobility, we introduce two quantities that measure organization-wide consistency between observed and potential organizational job transitions. 
Potential job transitions are created from random models constrained to preserve select features of the transitions. We generate three such models, one preserving labor supply and demand, another exclusively adding a condition to preserve the count of occupational transitions, and the last exclusively adding a condition to preserve the count of reuniting moves. Among them, the reuniting moves condition is by far the one leading to most consistency.

The main goal of this work is to provide empirical grounding to inform new approaches in the modeling of organizational job mobility, specifically identifying which system-wide mechanisms may have the greatest influences. In order to do this, we update the information used in structuring the system (the organization and its employees) to include an organization's team structure, social interaction among employees, and information that distinguishes each employee's work specialization (in our case, occupational specialization), all absent from extant models. In addition to dealing with the limiting simplifications of current models, the inclusion of these features connects VSMs and, generally, precise mobility models with other areas of organizational job mobility that have so far remained separate such as organizational social networks~\cite{podolny1997resources} and organizational team structure~\cite{purnam_book}. From a practical perspective, given the growth of more detailed data on organizations (e.g., see~\cite{Marler02012017}), we expect our framework to be applicable in practice, where it would indeed be possible to extract the same information we use here for other organizations. 

Our framework is built upon the detailed empirical analysis of the Army Acquisition Workforce (AAW), a large civilian and military organization that operates as part of the US Army. The AAW follows civilian employment practices generally consistent with those of the private sector, where individuals are free to pursue internal opportunities under the vacancy system. The data includes all the civilian AAW employees (over $70,000$) over a period of $9$ years. We construct the AAW teams based on this personnel information and a set of rules that define under which circumstances a group of employees form a team, how long that team exists, and when the team disbands.

\section{Methods}
\subsection{Data}\label{sec:methods-data}

The AAW is a United States Army organization consisting of uniformed service members and government civilians. In this study, we analyze AAW civilian personnel job positions, which are not based on military orders but on the organization's changing requirements. Therefore, the individuals analyzed are able to manage their career mobility within the AAW, contingent on opportunities or managerial decisions typical of private sector organizations. The data consists of monthly AAW personnel records from January 2012 to December 2020. In total, there are approximately $4$ million unique observations for over $70,000$ unique individuals in the data. An anonymized unique key links each employee record with their job title, an \textit{occupational series} as defined by the U.S. Office of Personnel Management~\cite{OPM}, the unique key of their supervisor, and the pay scale for the employee under the U.S. government's salary ladder. Over the period analyzed, the AAW ranged in size between approximately $35,000$ and $42,000$ government civilians.

\subsection{Generation of the time-dependent team structure of the AAW}\label{sec:methods-WU}
A methodological point we must address is how to determine the team composition of the organization. While this may appear to be a straightforward task, it is in fact not simple. There are both qualitative and practical difficulties that make the identification of organizational teams challenging. From a practical standpoint, most organizations maintain personnel information for purposes other than team function, among them budgets and payroll; this means that it is typical to know how much someone gets paid, but harder to find information about team composition. From a qualitative standpoint, the notion of a team implies the existence of a shared identity as well as a set of integrated tasks, something that requires a level of permanence and team stability.

To address the points raised above, here we proceed as follows. We begin by highlighting an advantage of our data, which is the linking of each employee to their supervisor, a feature that plays an important role in assigning individuals to teams. However, we note that it is not enough to create teams simply as those supervised by the same individual. This is because people (including supervisors) join and leave teams frequently for a variety of reasons and thus the composition of a team is rarely fixed~\cite{hom2017one}. Furthermore, management can actively affect team composition in several ways, adding to the complexity of defining a team. Thus, one requires criteria by which the temporal membership can be converted into well-defined teams over time. Assuming one can tackle this challenge, another emerges in that not every grouping of individuals in an organization necessarily constitutes a team. In particular, this is a source of ambiguity among individuals in middle management, who may share a supervisor but not collaborate on integrated tasks. 

\begin{figure}[H]
    \includegraphics[width=0.9\linewidth]{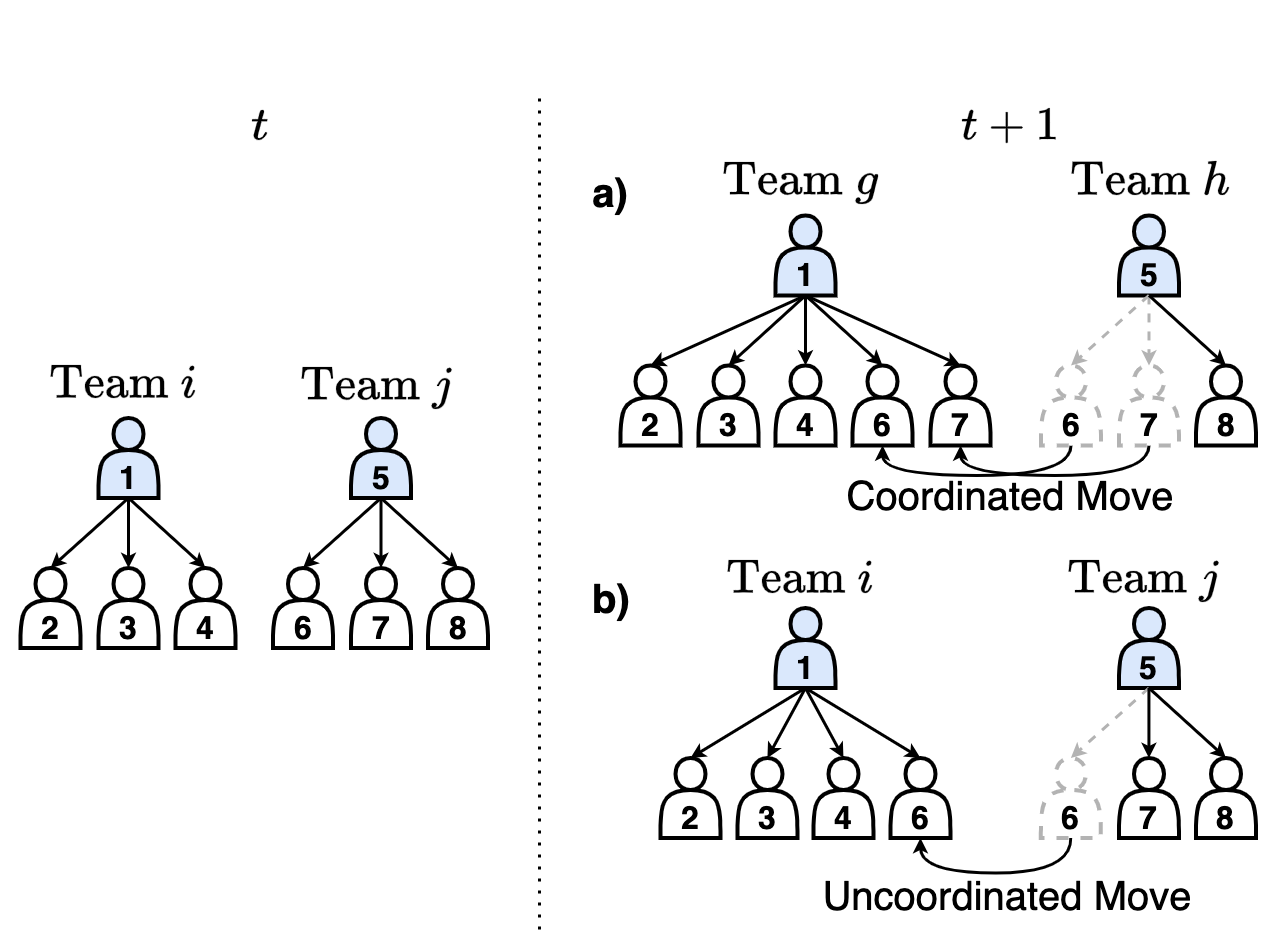}
    \caption{An illustration of coordinated and uncoordinated moves. This figure contains two teams, team $i$ and team $j$ during month $t$ (left of the dashed line) and $t+1$ (right of the dashed line). Between month $t$ and $t+1$ a coordinated move, a move of two or more people, could occur between team $i$ and $j$. Panel \textbf{a)} portrays this situation. When a coordinated move occurs both the losing team, $j$, and the gaining team, $i$ gain new identities, $i\rightarrow g$ and $j \rightarrow h$. Alternatively, an uncoordinated move, a move of only one person, between teams could occur. Panel \textbf{b)} portrays this situation. Teams do not gain a new identity when an uncoordinated move occurs between $t$ and $t+1$. Multiple uncoordinated moves between consecutive time steps do not change the identifies of the teams involved. However, if coordinated and uncoordinated moves occur, team identities do change.}
    \label{fig:coor_un_moves}
\end{figure}
Thus, we develop an empirical method based on our data to define teams. In our approach, each team has an identity (captured by a label such as $i$ or $j$) that persists in consecutive time steps $t$ and $t+1$ (one-month intervals) as long as two conditions are met between those steps: (i) from one time step to the next, there remain at least two members of the team (the permanence of the supervisor is not required) and (ii) team personnel are not part of any \textit{coordinated} moves (defined as two or more team members moving together between teams). These rules implicitly define the large majority of team births and deaths since, when a coordinated move occurs, in the next time step both the team that supplied the group of employees and the team that received them acquire new labels and the old team labels are eliminated (see Fig.~\ref{fig:coor_un_moves}(a)). These team births and deaths are akin to team \textit{mutations}. The logic behind these rules is that a move of multiple people simultaneously between teams is likely mandated by management, and signals an expectation for both the personnel-loosing and -gaining teams to operate differently than before the move. Furthermore, even if management is not expecting teams to operate differently after a coordinated move, the mere personnel churn is sufficiently large to induce significant change in team functioning~\cite{hale_turnover, hancock_turnover, turnover_analysis}. Far less frequently, teams can also be born or die from other mechanisms: full assembly or disassembly (when team members first unite or definitively disperse) or when a team member, except the team supervisor, ceases to be or becomes a supervisor of another team (see below).

Beyond coordinated moves, the majority of job changes can be characterized as \textit{uncoordinated} moves in which team members move among teams but not in groups of two or more in the same time step (see Fig.~\ref{fig:coor_un_moves}(b)). Multiple uncoordinated moves can occasionally occur, when several people depart or arrive at a team, but none are moving as a group. However, we note that when coordinated and uncoordinated moves occur simultaneously, the effect of coordinated moves takes precedence in that any teams involved in such moves change identities. By our definition, uncoordinated moves by themselves do not lead to the birth or death of a team. In contrast, the typically moderate personnel change associated with uncoordinated moves is common to many stable teams and in this case we assume team temporal continuity (i.e. no team label changes). In other words, a team experiencing only uncoordinated moves between time steps is considered to be a \textit{continuing team}, which is formally represented by maintaining its team label (and thus its team identity).

The definitions above lead to some team-specific properties of interest. First, we introduce $\ell_i$, the lifetime of team $i$, which corresponds to the time difference between the last and first months a team exists. A second property of interest is team size, labeled as $s_i$ for team $i$, which is defined as the number of members of team $i$. This property can potentially change over time and thus, in principle, it should be a function of $t$. In practice, our analyses deal with the small amount of change in ways described below. When discussing team lifetime in general (not relating to a specific team), we use $\ell$ without a subindex; the same is done in reference to size $s$. The general statistical properties of $\ell$ and $s$ of the teams generated by our approach can be found in Sec.~\ref{sec:SI-team-stats} and are consistent with those found in other empirical analyses of teams inside organizations~\cite{Horwitz, Bell}.

We note that, aside from the team supervisor, any team member that is herself/himself a supervisor of a group of employees disqualifies the team from our analysis. We call such teams \textit{non-terminal} in the sense that they are not located at the bottom of the organizational diagram. In this study, we concentrate on what we call \textit{terminal} teams because this avoids the ambiguity of middle management layers where people may be supervised by the same person but may not work collaboratively to perform specific tasks, hence not always constituting a true team.

\subsection{Network construction}
As shown in previous articles~\cite{labor_flows,inter_firm_labor_flows,linkedin}, job transitions among entities (e.g. firms in a large economy or large formal units within an organization) are well-represented by networks. Here, we also analyze the job transition problem using network concepts, although as we apply them our only concern are single connections among entities rather than more complex network structures such as long paths or network motifs. In our current approach, the nodes of the networks correspond to teams and the links between them represent uncoordinated job transitions (we use job transitions and network links interchangeably). The focus on uncoordinated moves stems from our interest in employee decision-making (and the fact that they are by far the majority of job moves). We concretely proceed as follows. 

Centered at time $t$ (a specific month), we take two periods, each of duration $\Delta t$, the first from $t-\Delta t$ to $t$ (denoted $\mathcal{T}_{<}$ for brevity) and the second from $t$ to $t+\Delta t$ (denoted $\mathcal{T}_{>}$). These definitions lead to a \textit{time window} of duration $2\Delta t$. Among all the teams that exist from $t-\Delta t$ to $t+\Delta t$, we focus on those that exist through the entire period; these are the continuing teams of the full time period $2\Delta t$. As implied from the discussion in Sec.~\ref{sec:methods-WU}, continuing teams have the same team label at least from $t-\Delta t$ to $t+\Delta t$. We then identify job transitions. Specifically,
during $\mathcal{T}_{<}$ a set of job transitions represented as links $\mathcal{E}_{<}$ are observed among continuing teams, with the subset $\vec{\mathcal{E}}^{(i)}_{<}\subseteq 
\mathcal{E}_{<}$ corresponding to the transitions starting at team $i$, and with the subset $\cev{\mathcal{E}}^{(i)}_{<}\subseteq \mathcal{E}_{<}$ the transitions ending at team $i$ (note the arrow is reversed). During $\mathcal{T}_{>}$, a set of transitions $\mathcal{E}_{>}$ occur across the teams, $\vec{\mathcal{E}}^{(i)}_{>}\subseteq\mathcal{E}_{>}$ out of team $i$ and $\cev{\mathcal{E}}^{(i)}_{>}\subseteq\mathcal{E}_{>}$ into of team $i$. 
Strictly speaking, all sets of links just defined are functions of both $t$ and $\Delta t$ (e.g. $\mathcal{E}_{<}(t,\Delta t$)) but we omit such specification to ease the notational burden.

\subsection{Modeling temporal consistency of job moves. Reuniting moves}
We first provide intuition about our approach to test for consistency. For us, consistency refers to how much a model, defined on the basis of preserving a set of chosen properties of organizational transitions, matches observed transitions. In other words, consistency is about how much a choice of properties is able to narrow down the space of possible observed transitions; a more consistent set of properties is synonymous with a closer guess for observed transitions. In what follows, we test consistency both forward and backwards in time, using information about one period of time to check how much it resembles another period of time. The intuition behind this is that, to introduce new approaches to VSMs or other mobility models, one must seek properties of job mobility that are informative about the time evolution of the system.

To define the concrete quantities used to test consistency, we begin by explaining the \textit{forward} time version and then define the \textit{reverse} version by a simple update of the quantities applied. 

Our check for \textit{forward consistency} of job moves is based on the recognition that in the time period $\mathcal{T}_{>}$, the observed transitions $\mathcal{E}_{>}$ are one of a set of possible outcomes. In other words, $\mathcal{E}_{>}$ is a single trial of a random variable of the job changes in $\mathcal{T}_{>}$. Here, we introduce three different models, labeled by the letter $m$, that allow us to generate trials of possible job transitions that could occur during $\mathcal{T}_{>}$ and compare these to the job transitions in the period $\mathcal{T}_{<}$. In the language of networks, all our models are \textit{uniform rewiring models} where the observed transitions $\mathcal{E}_{>}$ are reconnected under certain constraints, but are otherwise uniformly random. The random transitions are generated through Monte Carlo simulations: we create $1,000$ realizations for each value of $t$ in the range $[\Delta t+t_o,t_t-\Delta t]$, with $t_o=\text{January, 2012}$ and $t_f=\text{December, 2020}$. We use $\mathcal{S}_{m,n,>}$ to refer to the transitions randomly generated in the $n$th trial of model $m$ over the entire set of continuing teams, $\mathcal{S}^{(i,j)}_{m,n,>}$ for the specific transitions from node $i$ to $j$, $\vec{\mathcal{S}}^{(i)}_{m,n,>}$ for those starting at team $i$, and $\cev{\mathcal{S}}^{(i)}_{m,n,>}$ for those ending at $i$. To quantify forward consistency, we perform several comparisons involving $\mathcal{S}_{m,n,>}$, $\vec{\mathcal{S}}^{(i)}_{m,n,>}$, and $\cev{\mathcal{S}}^{(i)}_{m,n,>}$ against the job transitions that took place in the period before $t$, $\mathcal{T}_{<}$. In addition, since some teams have no transitions through the entire time window from $t-\Delta t$ to $t+\Delta t$, one version of consistency presented below is define so that it is able to quantitatively include the effect of these teams (see Sec.~\ref{sec:z}). 

\subsubsection{Models of job moves}\label{methods:models}
To generate our three models, we mainly focus on team properties but, in some cases, we include employee properties as well. The three model versions are as follows:
\begin{enumerate}
    \item \textit{Strength preserving model} (SP). Using the terminology of network science~\cite{newman2018networks}, the quantities $|\vec{\mathcal{E}}^{(i)}_{>}|$ and $|\cev{\mathcal{E}}^{(i)}_{>}|$ are, respectively, the out-strength and in-strength of node $i$ for the period $\mathcal{T}_{>}$. They represent the amount of flow out of and into each node. In the strength preserving model, we uniformly randomly assign job changes among continuing teams in such a way that $|\vec{\mathcal{S}}^{(i)}_{m,n,>}|=|\vec{\mathcal{E}}^{(i)}_{>}|$ and $|\cev{\mathcal{S}}^{(i)}_{m,n,>}|=|\cev{\mathcal{E}}^{(i)}_{>}|$. This model acts as a baseline for job change because the only information it relies on is supply and demand of labor at each team. 
    \item \textit{Occupation transition and strength preserving model} (OSP). To improve on the previous model, we add information about occupations which further restrict the possible transitions among teams and mimics the actual flow of expertise that occurs across the organization. Concretely, any employee that moves from team $i$ to team $j$, leaving a position with occupational series $u$ to take a position with occupational series $v$, adds a link to $\mathcal{E}_{>}$. Such links can be separated into categories that track occupational transitions such that we can write $\mathcal{E}^{(i,j)}_{>}(u,v)$ to represent the job transitions from $i$ to $j$ that take people from code $u$ to code $v$. In this OSP model, we randomly rewire transitions but preserve both node strengths and occupational transition counts, which is achieved by requiring that $\sum_j|\mathcal{S}^{(i,j)}_{m,n,>}(u,v)|=\sum_j|\mathcal{E}^{(i,j)}_{>}(u,v)|$ and $\sum_i|\mathcal{S}^{(i,j)}_{m,n,>}(u,v)|=\sum_i|\mathcal{E}^{(i,j)}_{>}(u,v)|$ for all $i$ and $j$ continuing teams.
    \item \textit{Coworker reuniting and strength preserving model} (RSP). In this model, we concentrate on the social content of job transitions. Based on our empirical findings (see Results), many job transitions reunite people that worked together in the past. To determine how much this effect contributes to consistent job transitions, we use information on each employee as well as the number of transitions between two teams that lead to coworker reunion. We call such job changes \textit{reuniting moves}. Let us label reuniting moves from continuing team $i$ to continuing team $j$ as $\mathcal{E}^{(i,j)}_{>}(r)$, while the non-reuniting moves as $\mathcal{E}^{(i,j)}_{>}(\neg r)$; similar symbols are applied to the random transitions. This model is defined by randomly rewiring transitions under the constraints $\sum_j|\mathcal{S}^{(i,j)}_{n,m,>}(r)|=\sum_j|\mathcal{E}^{(i,j)}_{>}(r)|$ , $\sum_j|\mathcal{S}^{(i,j)}_{n,m,>}(\neg r)|=\sum_j|\mathcal{E}^{(i,j)}_{>}(\neg r)|$, $\sum_i|\mathcal{S}^{(i,j)}_{n,m,>}(r)|=\sum_i|\mathcal{E}^{(i,j)}_{>}(r)|$ and $\sum_i|\mathcal{S}^{(i,j)}_{n,m,>}(\neg r)|=\sum_i|\mathcal{E}^{(i,j)}_{>}(\neg r)|$ for all $i$ and $j$ continuing teams. These constraints also preserve strength.
\end{enumerate}
As a matter of notation, when making reference to a specific model, we shall write $m$ to be equal to one of the model acronyms defined above (e.g. $m=\text{SP}$).

Before continuing with our presentation, we note that notions relevant to the concept of reuniting moves have been brought up before in a small number of articles, mostly at a qualitative level~\cite{Li,Ray}. They recognize that positive experiences by team members lead to a desire to remain connected to certain team members or to the team itself~\cite{Li}. Also, in cases where a team member departs, these positive experiences contribute to a greater willingness to rejoin the team in the future~\cite{Ray}. Reuniting moves are consistent with the well-established relevance of social interaction in the workplace~\cite{McEvily} and represent an expression of social capital~\cite{Coleman}. Furthermore, reuniting moves are consistent with the known phenomenon of job referrals~\cite{referrals-review}, which have been studied without being restricted to within-organization referrals. However, the authors are not aware of the concept of reuniting moves being explicitly defined nor measured until this study. 

\subsubsection{Consistency of moves team by team}\label{sec:z}
One approach to evaluate forward consistency of job moves is team by team, measuring how consistent job transitions into and out of each team may be during $\mathcal{T}_{>}$ in comparison to observed transitions during $\mathcal{T}_{<}$. As we now describe, this approach facilitates the inclusion of lack of job moves in the evaluation. 

To quantify team by team (i.e. node by node) forward consistency, we introduce the team-centric metric $z^{\text{forward}}_{m,n}$ for the $n$th realization of model $m$, which measures an aggregate score of consistency based on the nodes. The quantity is normalized as explained below. 

The steps to compute $z^{\text{forward}}_{m,n}$ are the following. We begin by determining for each node $i$ the proportion of repeated observed transitions between $\mathcal{T}_{<}$ and $\mathcal{T}_{>}$, given by the ratio  $(|\vec{\mathcal{E}}^{(i)}_{<}\cap\vec{\mathcal{E}}^{(i)}_{>}|+|\cev{\mathcal{E}}^{(i)}_{<}\cap\cev{\mathcal{E}}^{(i)}_{>}|)/(|\vec{\mathcal{E}}^{(i)}_{<}|+|\cev{\mathcal{E}}^{(i)}_{<}|)$. This ratio represents a score on the possible amount of exactly-repeated transitions into and out of team $i$, and is part of the normalization. Then, to score how well model $m$ performs, we produce the corresponding score for the $n$th simulated transitions $\vec{\mathcal{S}}^{(i)}_{m,n,>}$ and $\cev{\mathcal{S}}^{(i)}_{m,n,>}$ on node $i$, or $(|\vec{\mathcal{E}}^{(i)}_{<}\cap\vec{\mathcal{S}}^{(i)}_{m,n,>}|+|\cev{\mathcal{E}}^{(i)}_{<}\cap\cev{\mathcal{S}}^{(i)}_{m,n,>}|)/(|\vec{\mathcal{E}}^{(i)}_{<}|+|\cev{\mathcal{E}}^{(i)}_{<}|)$. Note that if $\vec{\mathcal{S}}^{(i)}_{m,n,>}=\vec{\mathcal{E}}^{(i)}_{>}$ and $\cev{\mathcal{S}}^{(i)}_{m,n,>}=\cev{\mathcal{E}}^{(i)}_{>}$, the score for each trial of random transitions is equal to the score for each set of real transitions. 

What happens when a team has no transitions during $\mathcal{T}_{<}$ and $\mathcal{T}_{>}$? In this case, conceptually speaking one would imagine that a model that correctly reproduces this behavior would also generate no transitions. In fact, this is guaranteed in all our models, as they all preserve node strength. However, the scoring approach proposed above in terms of ratios can generate $0/0$ terms. This requires us to introduce terms in the calculation of $z^{\text{forward}}_{m,n}$ that correctly handle these situations. 

The final step in constructing each realization $z^{\text{forward}}_{m,n}$ involves a way to compare the scores of consistency between observation and model, i.e. to normalize the score. To do this, we simply add all consistency scores of observed transitions team by team (including the no-transition teams), the scores of the modeled transitions in comparison to prior transitions (again team by team, including no-transition teams), and make a ratio between these two totaled scores. A ratio of $1$ is achieved when there is complete consistency between observations and model. We note that over the entirety of the realizations of random transitions, this normalization is typically bound from above by $1$. However, partly because the simulated transitions are about $\mathcal{T}_{>}$ while the observed transitions are about $\mathcal{T}_{<}$ (i.e. belong to different time periods), some isolated realization $n$ can lead to a $z^{\text{forward}}_{m,n}$ that is greater than $1$ but, as our results show, this is not an important effect. 

These considerations lead to the quantity 
\begin{equation}\label{eq:z_m}
    z^{\text{forward}}_{m,n}=\frac{\sum_{i\in\text{teams}}\left[\frac{|\vec{\mathcal{E}}^{(i)}_{<}\cap\vec{\mathcal{S}}^{(i)}_{m,n,>}|+|\cev{\mathcal{E}}^{(i)}_{<}\cap\cev{\mathcal{S}}^{(i)}_{m,n,>}|}{|\mathcal{E}^{(i)}_{<}|}+\delta_{|\mathcal{E}^{(i)}_{<}|+|\mathcal{E}^{(i)}_{>}|,0}\right]}
    {\sum_{i\in\text{teams}}\left[\frac{|\vec{\mathcal{E}}^{(i)}_{<}\cap\vec{\mathcal{E}}^{(i)}_{>}|+|\cev{\mathcal{E}}^{(i)}_{<}\cap\cev{\mathcal{E}}^{(i)}_{>}|}{|\mathcal{E}^{(i)}_{<}|}+\delta_{|\mathcal{E}^{(i)}_{<}|+|\mathcal{E}^{(i)}_{>}|,0}\right]},
\end{equation}
where we have introduced the shorthand notation $|\mathcal{E}^{(i)}_{<}|=|\vec{\mathcal{E}}^{(i)}_{<}|+|\cev{\mathcal{E}}^{(i)}_{<}|$ and $|\mathcal{E}^{(i)}_{>}|=|\vec{\mathcal{E}}^{(i)}_{>}|+|\cev{\mathcal{E}}^{(i)}_{>}|$ and where we treat any fraction $0/0$ to be equal to $0$. Note that $z^{\text{forward}}_{m,n}$ is a function of $t$ and $\Delta t$, which we omit for brevity.
In the case when the links of a model include all the links actually realized in $\mathcal{T}_{>}$, $z^{\text{forward}}_{m,n}$ becomes $1$; if model links bear no relation with past transitions, $z^{\text{forward}}_{m,n}$ becomes $0$. More broadly, we can understand $z^{\text{forward}}_{m,n}$ as a normalized proportional match, node by node and under model $m$, between possible future transitions during $\mathcal{T}_{>}$ and the real transitions. In the equation, the Kronecker $\delta_{c,d}$ is defined to be equal to $1$ when $c=d$ and $0$ otherwise, and is added to the numerator and denominator to score the teams with no flows in both $\mathcal{T}_{<}$ and $\mathcal{T}_{>}$. In Results, we also present a version of $z^{\text{forward}}_{m,n}$ that excludes the Kronecker deltas as these have a tendency to inflate $z^{\text{forward}}_{m,n}$. 

Defining the reverse consistency $z^{\text{reverse}}_{m,n}$ is done straightforwardly by swapping the choices of links between $\mathcal{T}_{>}$ and $\mathcal{T}_{<}$ or, simply put, reversing $<$ and $>$ in Eq.~\ref{eq:z_m}. This leads to the quantity
\begin{equation}\label{eq:z_r_m}
    z^{\text{reverse}}_{m,n}=\frac{\sum_{i\in\text{teams}}\left[\frac{|\vec{\mathcal{E}}^{(i)}_{>}\cap\vec{\mathcal{S}}^{(i)}_{m,n,<}|+|\cev{\mathcal{E}}^{(i)}_{>}\cap\cev{\mathcal{S}}^{(i)}_{m,n,<}|}{|\mathcal{E}^{(i)}_{>}|}+\delta_{|\mathcal{E}^{(i)}_{<}|+|\mathcal{E}^{(i)}_{>}|,0}\right]}
    {\sum_{i\in\text{teams}}\left[\frac{|\vec{\mathcal{E}}^{(i)}_{<}\cap\vec{\mathcal{E}}^{(i)}_{>}|+|\cev{\mathcal{E}}^{(i)}_{<}\cap\cev{\mathcal{E}}^{(i)}_{>}|}{|\mathcal{E}^{(i)}_{>}|}+\delta_{|\mathcal{E}^{(i)}_{<}|+|\mathcal{E}^{(i)}_{>}|,0}\right]},
\end{equation}
where some terms are invariant under the reversal. It should be noted that the definition of $z^{\text{reverse}}_{m,n}$ generates an information gap in all the models but it is specially important in the $\text{RSP}$ model because reuniting information for the random links in $\mathcal{S}_{m,n,<}$ relies on work histories up to $t-\Delta t$, which are then compared to job transitions in the interval $t$ to $t+\Delta t$, leaving  a gap of $\Delta t$ units of time. This gap is not present for $z^{\text{forward}}_{m,n}$ which compares adjacent time intervals. We discuss this effect in more detail in Sec.~\ref{sec:yz-results}.

As a shorthand for both forward and reverse team consistency, we use the symbol $z_{m,n}$ without the super-index.

\subsubsection{System-wide consistency of job moves over the entire AAW}
To provide a complementary assessment of how consistent each model $m$ is to observations, we introduce the system-wide consistency $y_{m,n}$ that takes into account the overall link structure of the networks in the periods $\mathcal{T}_{<}$ and $\mathcal{T}_{>}$. In its forward version, $y^{\text{forward}}_{m,n}$, it is symbolically defined as
\begin{equation}\label{eq:ymn}
    y^{\text{forward}}_{m,n}=\frac{|\mathcal{E}_{<}\cap\mathcal{S}_{m,n,>}|}{|\mathcal{E}_{<}|},
\end{equation}
which is normalized and where $1$ can be reached only if the transitions $\mathcal{S}_{m,n,>}$ match $\mathcal{E}_{<}$ exactly. In general, $y^{\text{forward}}_{m,n}$ should be interpreted as the normalized match score given by model $m$ of possible future transitions in comparison to observed transitions. The quantity $y^{\text{forward}}_{m,n}$ does not count lack of transitions, which illustrates the complementarity between $y^{\text{forward}}_{m,n}$ and $z^{\text{forward}}_{m,n}$. As $z^{\text{forward}}_{m,n}$, $y^{\text{forward}}_{m,n}$ depends on $t$ and $\Delta t$. 

In the same way as for the team by team consistency, we also define the reverse system-wide consistency by swapping $<$ and $>$, leading to
\begin{equation}\label{eq:y_r_mn}
    y^{\text{reverse}}_{m,n}=\frac{|\mathcal{E}_{>}\cap\mathcal{S}_{m,n,<}|}{|\mathcal{E}_{>}|}.
\end{equation}

As a shorthand for both forward and reverse system consistency, we use the symbol $y_{m,n}$ without the super-index.

\subsection{Reuniting probabilities}\label{sec:ps-pm}
Given the novelty of the \textit{reuniting moves} definition, it is useful to develop intuition about the factors affecting them. Therefore, we define the concept of \textit{reuniting probability} and calculate its relation to several variables. 

We concentrate on two variables that may affect reuniting probability. First we check if team sizes are related to reuniting probability. This can be justified intuitively by the fact that an individual has a finite capacity to develop strong relationships~\cite{dunbar,BERNARD1973145,signatures}. Furthermore, teams of sizes above the typical employee's relationship-forming capacity may lead to forming weaker social bonds, thus a lower desire to reunite with members of such teams. We denote by $p(s)$ the reuniting probability as a function of team size $s$. To calculate $p(s)$, we add all reuniting moves that lead to employees reuniting in a team given that when they stopped working together in the past, that team had size $s$, and divide this number by the total number of reuniting moves.

Second, we study the dependence of reuniting moves on the duration of time that two people spent working together in the past. Interest in this variable is justified by the fact that a longer period of time of collaboration is likely to lead to the formation or stronger bonds, increasing people's familiarity with one another. Labeling as $\sigma$ the number of months two employees spent working together in the last team where they coincided, reuniting probability $p(\sigma)$ is calculated as the ratio of reuniting moves that lead to the re-encounter of people that have previously worked together $\sigma$ months divided by all reuniting moves.

Because time of service within the AAW affects the samples of employees that could have worked together for a certain amount of time, we create a subsample of the data, which we call $\mathcal{A}_{\Delta t}$, satisfying the condition that all employees in the sample have entered the AAW after the beginning of our data (thus, they would have joined the organization after January 2012) and have spent a minimum of $\Delta t$ months in the organization. We take $\Delta t=60$ (i.e. $5$ years), which leads to a sample of 3,333 AAW members who participated in 1,897 uncoordinated moves resulting in 625 (32.9\%) uncoordinated reuniting moves. 

Finally, to corroborate the trends founds for $p(s)$ and $p(\sigma)$, we created a logistic regression model. We apply it as follows: if an employee re-encountered a former coworker, that observation $n$ is considered a \textit{success} and encoded as $R_n = 1$; a move that does not lead to a re-encounter with a former coworker is considered a \textit{failure} and encoded as $R_n = 0$. We then model $R_n=1$ as a function of minimum team size $s_n$, defined as the smallest team size shared with a former team member, and the number of months worked together prior to the re-encounter, $\sigma_n$, as the explanatory variables. The model is specified by
\begin{equation}
    \label{eq:log_reg_re_encounter}
    Pr(R_n = 1|s_n,\sigma_n) = \frac{\exp{\left(\beta_0 + \beta_1 s_n + \beta_2 \sigma_n\right)}}{1 + \exp{\left(\beta_0 + \beta_1 s_n + \beta_2 \sigma_n\right)}}.
\end{equation}
In total there are 37,596 observations used in this model.

\section{Results}\label{sec:results}
We begin by providing statistical information about continuing teams, which offers context about their relevance to the overall organization. We then proceed to study consistency metrics of job transitions, showing evidence that social interaction plays a dominant role affecting those transitions. Finally, we show how team size and duration of prior collaboration increases the chances of prior coworkers to reunite.

\subsection{Continuing teams}

We first show that continuing teams represent an important component of the entire organization, particularly the employees. For this purpose, we measure the numbers of continuing and non-continuing teams there are between the start of 2012 and the end of 2020 that last at least $1$ year ($\Delta t = 6$), as well as the person-months of people in those continuing teams. By averaging month by month over the $9$ years of the data, the AAW has $\approx 2,533.55$ teams (covering $\approx 217,213.31$ people-months) that maintain their identity at least $1$ year and $4,382.58$ teams (covering $148,272.11$ people-months) which undergo a change of identity in the equivalent time window. It is noteworthy that although the number of teams that maintain their identity is smaller than those that change, the number of people-months for the stable teams is considerably greater than for the changing teams. For context, the average number of person-months in a 1-year window for \textit{all} AAW employees is equal to $429,392.63$, which means that just over $50\%$ of all person-hours occur in continuing teams. The main conclusion of this analysis is that a large fraction of the organization is captured in the behavior of continuing teams and their personnel. In the Supplementary Information, we show the numbers of teams, persons, and person-hours for other $\Delta t$ to provide a broader perspective for these numbers. 

Our choice of a $1$-year time window mainly stems from the measurement of team member tenure (the length of time an individual spends in a team). Calculating the quantity over all the unique teams identified through the $9$-year range of our data, which leads to a total of 31,697 unique teams, we find that average team member tenure is $\approx 12.14$ months. A second argument providing justification to use a $1$ year window comes from prior research that shows that, in other employment contexts, yearly periodicity in job tenure is common~\cite{Farber}. 

\subsection{Consistency of job moves based on different models}\label{sec:yz-results}
\begin{figure}
    \centering
    \includegraphics[width=0.8\linewidth]{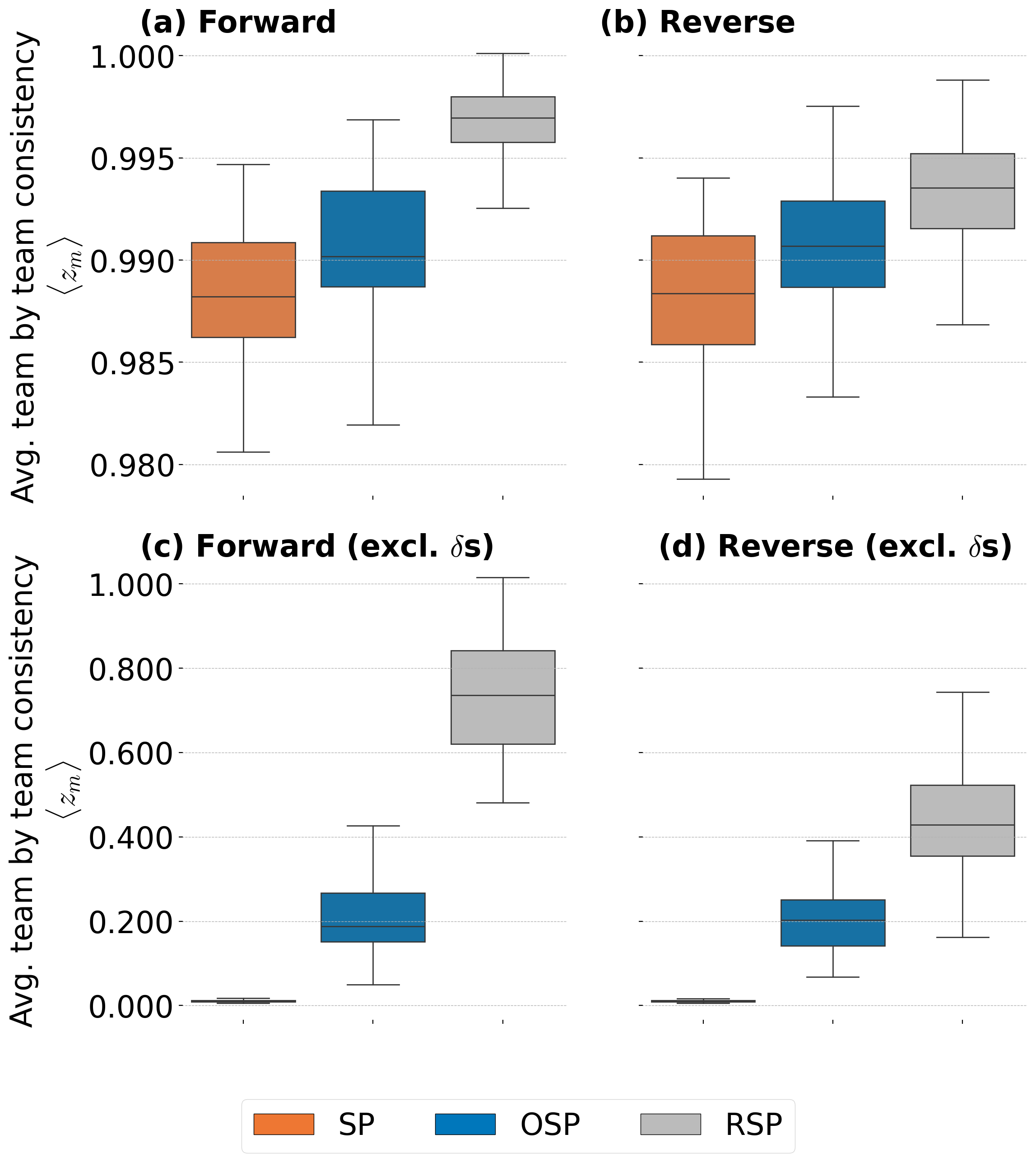}
    \caption{Box plots for team by team forward and reverse consistency over all models $m$. Each point of a box plot is composed of the averaged results of 1,000 realizations for each model $m$. Each box plot contains $96$ points, one for each possible $1$-year window during the time span of the data. Panel \textbf{(a)} contains the average of $z^{\text{forward}}_{m,n}$ calculated using Eq.~\ref{eq:z_m} and panel \textbf{(b)} the corresponding average for $z^{\text{reverse}}_{m,n}$ from Eq.~\ref{eq:z_r_m}. Panel \textbf{(c)} contains the average of $z^{\text{forward}}_{m,n}$ using Eq.~\ref{eq:z_m} without $\delta_{|\mathcal{E}^{(i)}_{<}|+|\mathcal{E}^{(i)}_{>}|,0}$ in the numerator and denominator and \textbf{(d)} contains the equivalent average of $z^{\text{reverse}}_{m,n}$. The results show that $m=\text{RSP}$ leads to the largest consistency in transitions.}
    \label{fig:zm}
\end{figure}

To compare the different possible features that may drive job moves, we present Fig.~\ref{fig:zm}, which shows the statistics of $z_{m,n}$ under the three models presented in Sec.~\ref{methods:models}. Figs.~\ref{fig:zm}(a) and~\ref{fig:zm}(c) show forward consistency $z^{\text{forward}}_{m,n}$ and Figs.~\ref{fig:zm}(b) and~\ref{fig:zm}(d) reverse consistency $z^{\text{reverse}}_{m,n}$.
Each point in each box plots corresponds to the average $\langle z_{m}(t)\rangle =\sum_{n=1}^{1,000}z_{m,n}/1,000$ over Monte Carlo realizations. Each box plot (one for every $m$) consists of $t_f-t_o-2\Delta t=96$ points. 
First, we apply Eq.~\ref{eq:z_m} leading to plot Fig.~\ref{fig:zm}(a). The performance of all three models is excellent, where any $m$ leads to values of $z^{\text{forward}}_{m,n}$ almost equal to $1$. A second important observation is that, among the three models, $m=\text{RSP}$ performs the best. However, the value of $z^{\text{forward}}_{m,n}$ close to $1$ should be interpreted with care, as it is driven by the fact that many continuing teams do not have transitions in either $\mathcal{T}_{<}$ or $\mathcal{T}_{>}$ (on average $55.3\%$ for $\Delta t= 6$ months) which leads to an abundance of terms equal to $1$ in both the numerator and denominator in Eq.~\ref{eq:z_m} (this is the contribution of the Kronecker deltas). Since it is reasonable to consider that there is a difference between teams with transitions or without transitions, we create a version of $z^{\text{forward}}_{m,n}$ where we only use teams that had transitions in both $\mathcal{T}_{<}$ and $\mathcal{T}_{>}$, i.e. ignoring the Kronecker deltas in Eq.~\ref{eq:z_m}, which provides a complementary assessment of each model $m$. These results are presented in Fig.~\ref{fig:zm}(c). In this case, performance of the models becomes clearly distinguishable. Model $m=\text{SP}$ performs poorly, $m=\text{OSP}$ performs better but still rather poorly, but model $m=\text{RSP}$ is considerably better, with values of $z_{m,n}$ including interquartile ranges going from just below $0.6$ to just over $0.8$. 

Reverse team by team consistency $z^{\text{reverse}}_{m,n}$ (Eq.~\ref{eq:z_r_m}) performs the same as $z^{\text{forward}}_{m,n}$ qualitatively in that the consistency is greatest for $\text{RSP}$, intermediate for $\text{OSP}$, and least for $\text{SP}$ (see Fig.~\ref{fig:zm}, panels b and d). The only caveat is that $m=\text{RSP}$ shows a drop with respect to forward consistency. As explained in Sec.~\ref{sec:z}, reverse consistency is defined with a gap in information, and this affects knowledge about reuniting moves in that it ignores a period of employees' job histories between $t-\Delta t$ and $t$. Although this gap is also present for the other models, occupational transition information and supply and demand information is aggregate over the entire system, which makes it more stable and thus less affected by the time gap in information.

\begin{figure}
    \centering
    \includegraphics[width=0.8\linewidth]{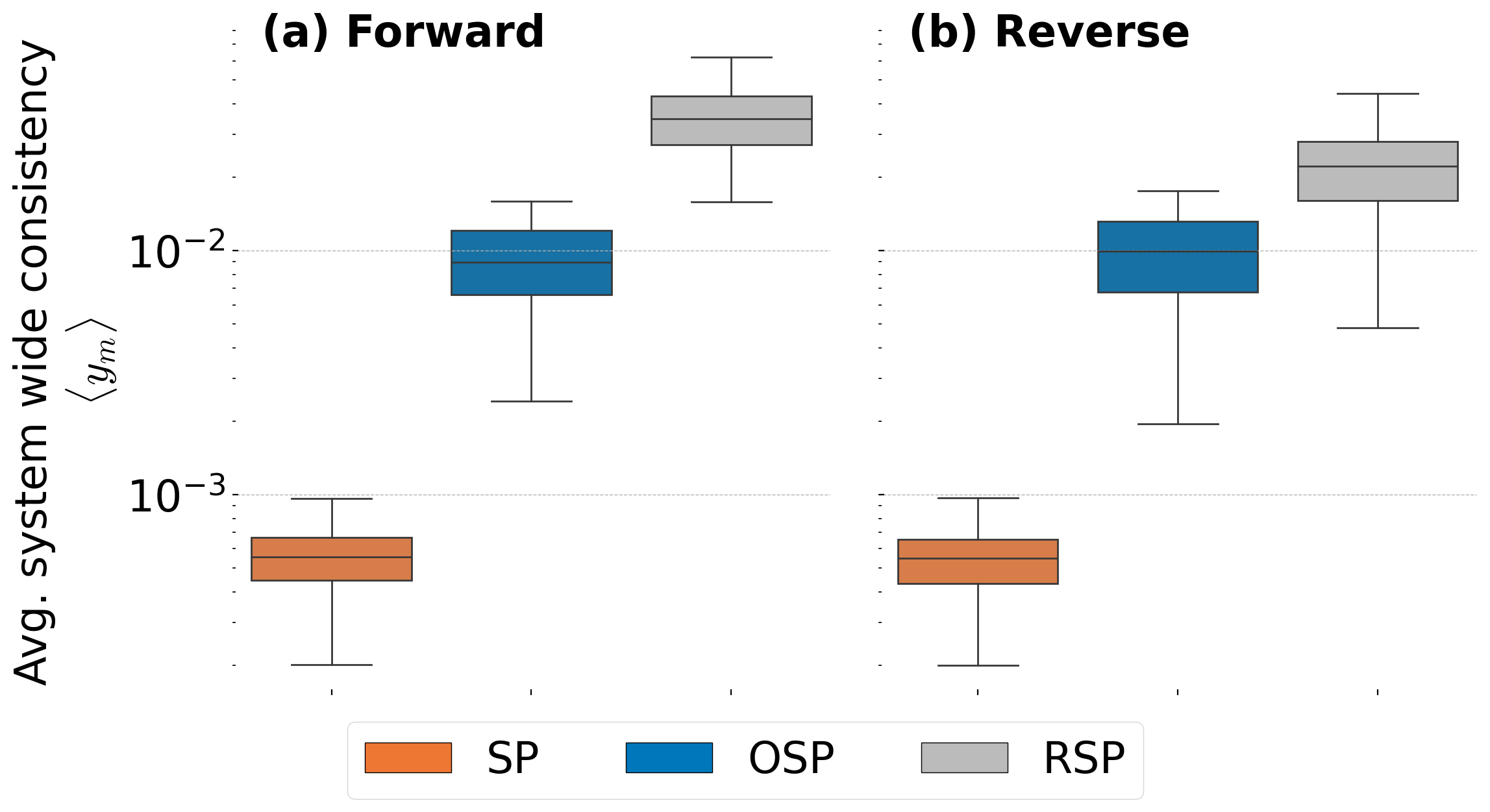}
    \caption{Box plots for system-wide forward and reverse consistency. Each point of a box plot is composed of the average of the 1,000 realizations of $y_{m,n}$ for each model $m$. Each box plot contains $96$ points, one for each possible $1$-year window during the time span of the data. Panel \textbf{(a)} is constructed from forward consistency $y^{\text{forward}}_{m,n}$ and panel \textbf{(b)} from reverse consistency $y^{\text{reverse}}_{m,n}$. It is clear that $m=\text{RSP}$ provides a better consistency of possible transitions.}
    \label{fig:ym}
\end{figure}
To study consistency for the entire system, we now turn to $y_{m,n}$, given by Eqs.~\ref{eq:ymn} and~\ref{eq:y_r_mn}. Here, scoring gives equal weight to each job transition across the system, in contrast to $z_{m,n}$ where consistent transitions among teams that have many job changes are weighted less than among teams with few job changes; in other words, $z_{m,n}$ is biased towards teams with few transitions. Thus, in Fig.~\ref{fig:ym} we show $y_{m,n}$ for the three models in both forward (Fig.~\ref{fig:ym}(a)) and reverse (Fig.~\ref{fig:ym}(b)) versions. As in the case of $z_{m,n}$, $m=\text{RSP}$ is the top performer, $m=\text{OSP}$ achieving second best performance and, finally, $m=\text{SP}$ performing worst. Similar to the case of $z_{m,n}$, each point of each box plot corresponds to an average over Monte Carlo realizations of one time point, i.e. $\langle y_m(t)\rangle=\sum_{n=1}^{1,000}y_{m,n}(t)/1,000$. Here, we again see a slight drop for $m=\text{RSP}$ between the forward and reverse versions due to the information gap from $t-\Delta t$ to $t$.

It is useful to take stock of the results above. Since consistency measures how similar possible and observed job transitions are to each other under the conditions imposed by a given $m$, an $m$ with greater consistency indicates that it better matches possible and observed job moves. Thus, the results shown in Figs.~\ref{fig:zm} and~\ref{fig:ym} indicate that a better way to understand job transitions across an organization is to know who has worked with whom. This outperforms a simple supply and demand or an occupational skills understanding, indicating that the social connections people form among each other can be more indicative of the internal change dynamics than seemingly more critical variables such as people's occupational specialization. 

\subsection{Reuniting is more important than occupation}
\begin{figure}
    \centering
    \includegraphics[width=0.8\linewidth]{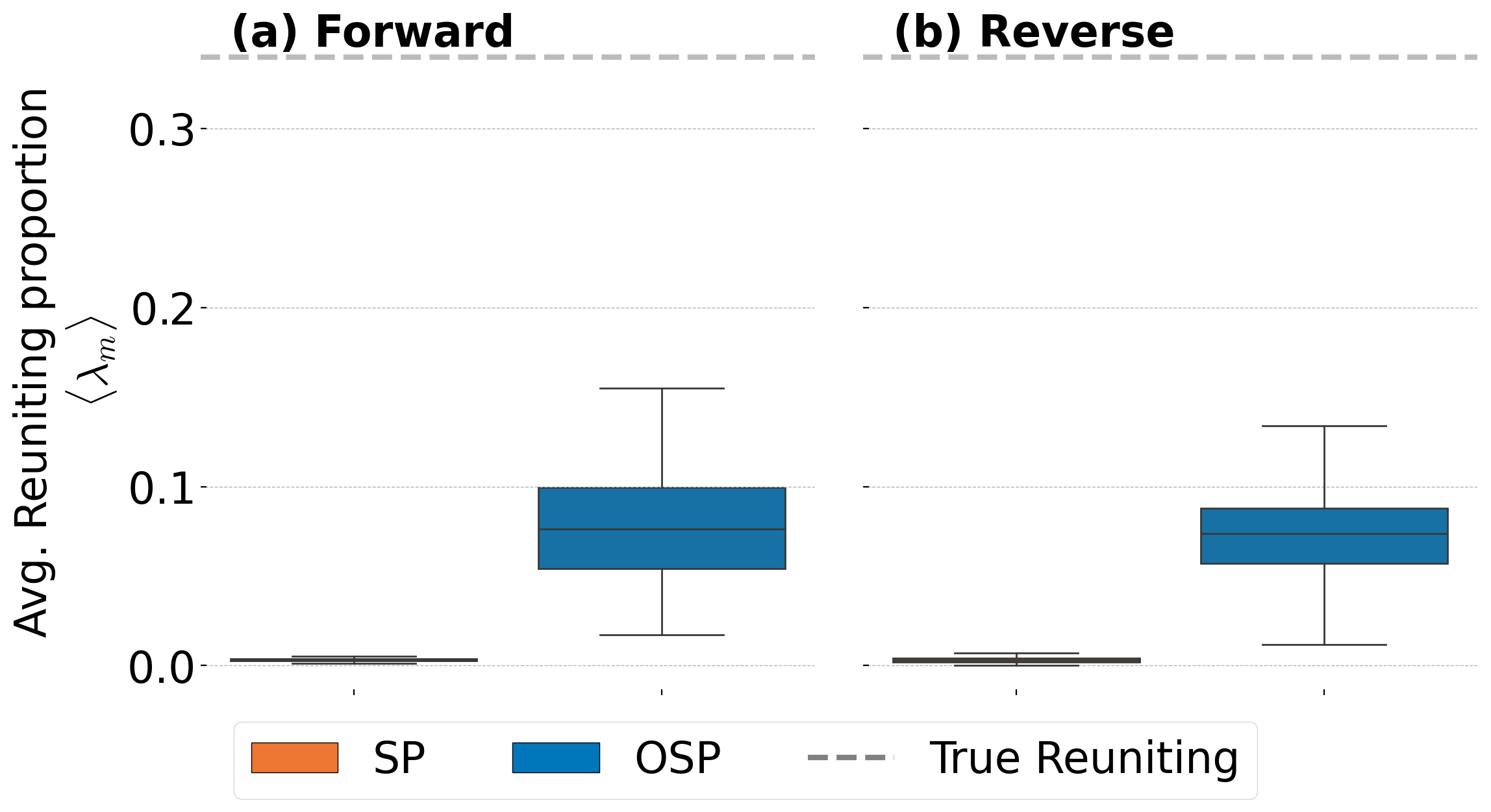}
    \caption{Box plots for the ratio of observed versus random reuniting moves from the $\text{SP}$ and $\text{OSP}$ models. Each point in each box plot is composed of the average of 1,000 realizations of $\lambda_{m,n}$ for the $\text{SP}$ and $\text{OSP}$ models. Each box plot contains $96$ points, one for each possible $1$-year window during the time span of the data. Panel \textbf{(a)} shows the ratio between observed and random job moves during $\mathcal{T}_{>}$ and panel \textbf{(b)} the equivalent during $\mathcal{T}_{<}$. The $\text{OSP}$ model produces more reuniting moves than $\text{SP}$, although these are only about $23\%$ of the actual number of reuniting moves (see Tab.~\ref{tab:following-proportions}). The dashed line above both box plots shows the value of reuniting moves directly measured in the data ($\approx 0.34$-- see Tab.~\ref{tab:following-proportions}).}
    \label{fig:per_follow}
\end{figure}
An additional analysis gives further evidence that the consistency in job mobility is more strongly related to reuniting than occupational expertise. In Fig.~\ref{fig:per_follow}, we measure how often moves dictated by $m=\text{SP}$ or $\text{OSP}$ lead to a reuniting move. By design, these two models do not require that reuniting move counts be preserved, so moves that do lead to workers reuniting are driven solely by the conditions the models $\text{SP}$ or $\text{OSP}$ impose. 

We define $\lambda^{\text{forward}}_{m,n}$, which depends on $t$ and $\Delta t$, as the ratio between the number of reuniting moves that occur among the transitions $\mathcal{S}_{m,n,>}$ in realization $n$ of model $m$ divided by the actual reuniting moves that occur in $\mathcal{E}_{>}$. We equivalently define $\lambda^{\text{reverse}}_{m,n}$ using reuniting moves in $\mathcal{S}_{m,n,<}$ and $\mathcal{E}_{<}$. Further, we determine an average $\langle\lambda_m(t)\rangle=\sum_{n=1}^{1,000}\lambda_{m,n}(t)/1,000$ for each possible $t$. The results for $m=\text{SP}$ (orange) and $m=\text{OSP}$ are displayed in Fig.~\ref{fig:per_follow}. The figure clearly show that $m=\text{SP}$ leads to the least number of reuniting moves. The $m=\text{OSP}$ model exhibits more reuniting moves, but still typically about $23\%$ of what is actually observed (less than $8\%$ of the totality of moves under $\text{OSP}$ are reuniting compared to $34\%$ of the observed moves--see Tab.~\ref{tab:following-proportions}).

\begin{table}[tb]
    \centering
    \newcolumntype{C}{>{\raggedright\arraybackslash}X}
    \begin{tabularx}{0.7\linewidth}{Crrr}
        \toprule
        Direction & SP & OSP & Measured reuniting \\
        \midrule
        Forward & $0.0032$ & $0.078$ & \multirow{2}{*}{$0.34$}\\
        Reverse & $0.0031$ & $0.075$ & \\
        \bottomrule
    \end{tabularx}
    \caption{Averages of $\lambda_{m,n}$ for each of the box plots presented in Fig.~\ref{fig:per_follow}. These averages are compared to the measured proportion of reuniting moves, averaged over the same time windows. It is clear, regardless of modeling direction, that both the $\text{SP}$ and $\text{OSP}$ perform quite poorly in terms of maintaining the observed proportion of reuniting moves in the AAW. For $m=\text{SP}$, comparing its ratio to the fraction of reuniting moves, we find $z$-scores of -10.73 (forward) and -11.00 (reverse); for $m=\text{OSP}$ models, we find $z$-scores of -7.68 (forward) and -8.01 (reverse). This indicates that reuniting moves are not merely a consequence of satisfying other factors influencing transitions, but are in themselves a dominant factor in transition choice.}
    \label{tab:following-proportions}
\end{table}

In order to summarize the information contained in Fig.~\ref{fig:per_follow}, we calculate the averages of the values in each of the box plots in Fig.~\ref{fig:per_follow}. These numbers are presented in Tab.~\ref{tab:following-proportions} for both forward and reverse random realizations of the $\text{SP}$ and $\text{OSP}$ models. The results are highly significant statistically as can be appreciated from the $z$-scores reported in the table caption. The amount of reuniting occurring across the organization is much larger than what the $\text{SP}$ and $\text{OSP}$ models indicate.

Interpreting the results in Figs.~\ref{fig:zm}, \ref{fig:ym}, and~\ref{fig:per_follow} and Tab.~\ref{tab:following-proportions}, we observe that the $m=\text{RSP}$ model is more consistent than the other models and, in addition, the other models have fewer reuniting moves than what is observed. This means that any plausible updates to the way in which mobility models are defined would do well to take into account the tendency for people to reunite with previous coworkers, i.e. social interaction. 

\subsection{Factors influencing reuniting moves}
As we argue above, social bonds are expected to be solidified with interaction duration but can be diluted in large groups. To verify this, we perform two analyses with two different employee samples, the entire workforce (which we call $\mathcal{A}$) and the sample $\mathcal{A}_{\Delta t=60\ \text{months}}$ as described above (see Sec.~\ref{sec:ps-pm}).

\begin{figure}[H]
    \centering
    \includegraphics[width=\textwidth]{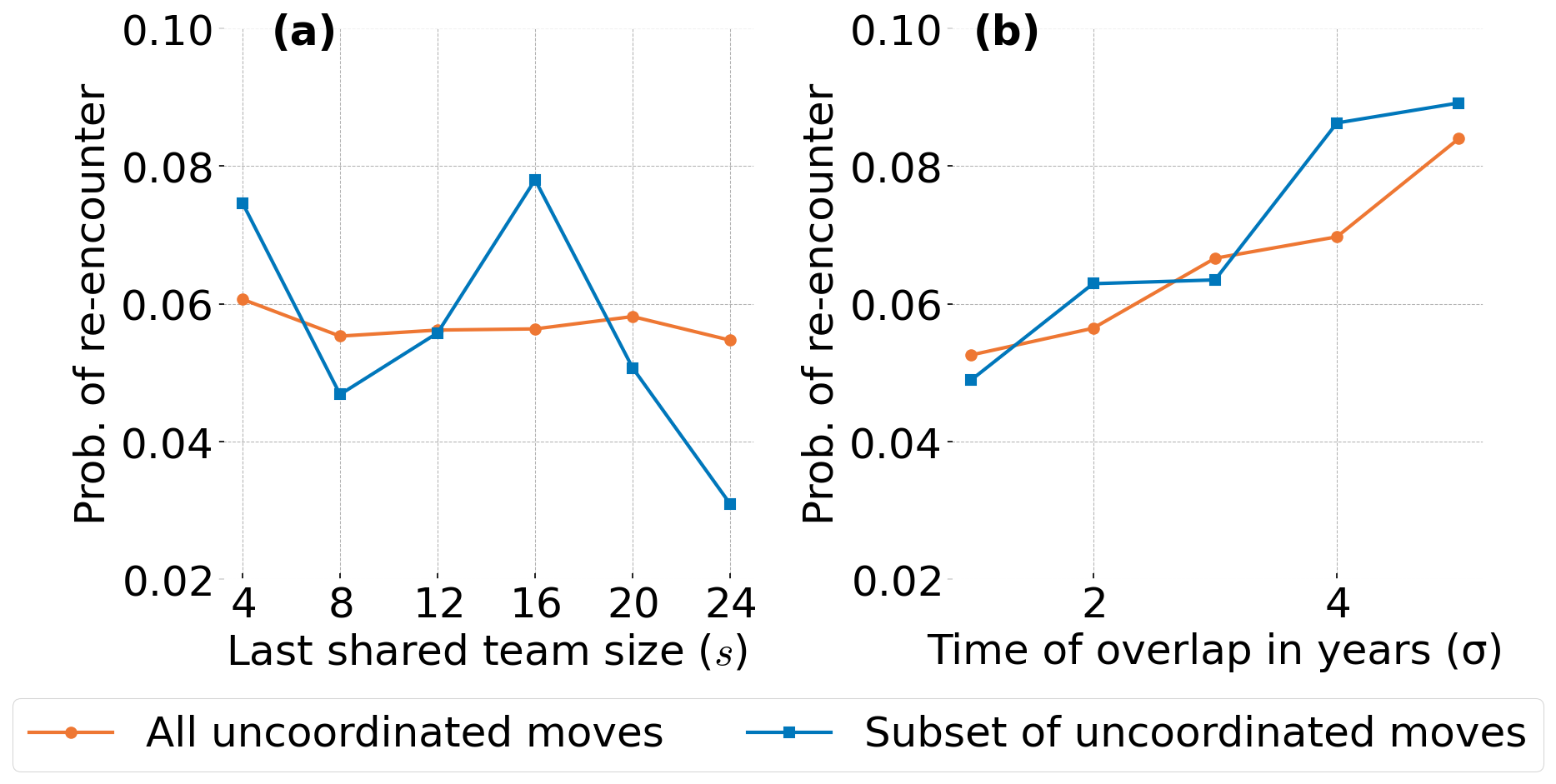}
    \caption{Reuniting probability as a function of previous co-working arrangement. Panel \textbf{(a)} shows the probability $p(s)$ for two past coworkers that last worked together in a team of size $s$ to get reunited again. As the size $s$ increases, on average, there is a decrease in the probability of a reuniting move. Team size $s$ is counted in bins of size 4. Panel \textbf{(b)} shows the probability of reuniting with a former team member given that they worked together in their last team in common $\sigma$ months in the past (months binned up to the next integer number of years). As the number of years working together on a team increases, the probability of reuniting increases. Both relationships are supported by the multivariate logistic regression model found in Table~\ref{tab:log_reg_results}. Observations from set $\mathcal{A}$ are represented by \protect\tikz[baseline=-0.5ex]{
    \protect\draw[wu_color, thick] (0,0) -- (.5,0);
    \protect\fill[wu_color] (0.25,0) circle (2pt);
} and observations from set $\mathcal{A}_{\Delta T}$ are represented by \protect\tikz[baseline=-0.5ex]{
    \protect\draw[uic_color, thick] (0,0) -- (.5,0);
    \protect\filldraw[uic_color] (0.20,-0.05) rectangle (0.3,0.05);
}.}
    \label{fig:reg_figures}
\end{figure}

In Fig.~\ref{fig:reg_figures}, we show the probabilities $p(s)$ and $p(\sigma)$ as described in Sec.~\ref{sec:ps-pm}. For $p(s)$, we count all job transitions where two individuals reunite after having last worked together in a team that had size $s$ in the last month before the employees stopped working together. This count is divided by all reuniting moves among teams of any size. The values of $s$ are binned as indicated in the figure caption. As Fig.~\ref{fig:reg_figures}(a) shows, the trend is weakly decaying for both $\mathcal{A}$ and $\mathcal{A}_{\Delta t=60\ \text{months}}$ although the latter sample shows an initial convexity ranging between approximately $5\%$ and $7\%$ probability for team sizes up to about a dozen individuals, followed by a more precipitous decrease to approximately a $3\%$ probability. This suggests that while the overall dependence decays with team size, there may be more detailed dependencies at play.

The plot associating reuniting probability to the time two individuals worked together is shown in Fig.~\ref{fig:reg_figures}(b) where times have been binned to yearly intervals (see caption). As with the calculation of the previous probability, to generate the probability we count all those reuniting moves that reunite employees that previously worked together for a period of larger than $\sigma-1$ up to $\sigma$, and divide by all reuniting moves. In this case, $p(\sigma)$ is clearly increasing and also consistent between samples $\mathcal{A}$ and $\mathcal{A}_{\Delta t=60\ \text{months}}$. 

In an effort to further corroborate our intuition regarding the effects of $s$ and $\sigma$ on the inclination of people to follow others, for our second analysis we built the multivariate logistic regression model ($n=37,596$) specified by Eq.~\ref{eq:log_reg_re_encounter}, to measure the impact of these two factors on the probability of a re-encounter (the results are found in Tab.~\ref{tab:log_reg_results}). Both factors, team size ($s$) and time working together ($\sigma$), have a statistically significant impact on the probability of re-encounter ($R=1$). For every increase in ten members of a team, the odds of a re-encounter among the sample $\mathcal{A}_{\Delta T=60\ \text{months}}$ decreases by a factor of $0.87$ and for every increase of one year working with a member, the odds of a re-encounter increases by a factor of $1.04$. 

\begin{table}[tb]
    \centering
    \newcolumntype{C}{>{\raggedright\arraybackslash}X}
    \begin{tabularx}{0.9\linewidth}{Crrr}
        \toprule
        Feature & Coefficient & p-value & 95 \% CI \\
        \midrule
        Intercept &	-2.4547	& $<$ 0.001 &  (-2.536, -2.373)\\
        Minimum shared team size ($s$) & -0.0141 & $<$ 0.001 & (-0.017, -0.011) \\
        Number months overlapped ($\sigma$) &	0.0033 & 0.050 & ($1.45 \times 10^{-6}$, 0.007)\\
        \bottomrule
    \end{tabularx}
    \caption{Logistic regression model results for social capital influence on an individual's career transition based on the sample $\mathcal{A}_{\Delta t=60\ \text{months}}$.\label{tab:log_reg_results}}
\end{table}

\section{Discussion}\label{sec:discussion}
In this article, we introduce a new framework to study job mobility based on decomposing an organization into teams. Concentrating on large data covering an extensive period of time for the Army Acquisition Workforce, we focus on job transitions among \textit{terminal continuing teams} given a time window, which constitute an important fraction of all organizational teams and carry the majority of the people-months of the personnel in the organization. The identification of organizational teams allows us to determine social relationships between individuals and how these may impact subsequent employee organizational job moves. We introduce three different models that dictate how people change jobs among terminal continuing teams and check how each model performs in terms of generating job changes that are consistent with past job changes. The models used preserve (i) labor supply and demand ($\text{SP}$), (ii) supply and demand specialized to occupations ($\text{OSP}$), or (iii) supply and demand that also preserves preferences for former colleagues to reunite ($\text{RSP}$). We find that the $\text{RSP}$ model shows the largest amount of consistency with job transitions. We also find that social preferences to reunite with colleagues are strongest for those that previously worked in smaller teams and for longer periods of time. 

The numbers of reuniting moves predicted by the $\text{SP}$ and $\text{OSP}$ models are well below the number of observed reuniting moves, making evident that these models do not accurately reflect the \textit{quantitatively dominant} tendencies affecting people's mobility. Therefore, we find support for the conclusion that the social aspect plays an oversized role in job mobility. 
We also note that, from the perspective of an employee performing a job move, the number of other teams they can move to and reunite with a past coworker is fairly limited, more so than the number of teams they can join to continue to work in a job with the same occupational specialization, for instance. This explains why the $\text{RSP}$ model performs better in consistency metrics than the other models. This also suggests that the addition of a reuniting condition to VSMs or other mobility models may be a way to design high-performing models. 

It is well-known that social ties constitute channels for information sharing~\cite{strength_of_ties}, brokerage~\cite{Burt}, and even company operation~\cite{McEvily,company-chart}. Furthermore, it has been shown that organizational social networks play a role in the interplay between workplace performance and mobility~\cite{podolny1997resources}. Here, by introducing the concept of job reunions, we highlight how social interactions directly influence job changes to the point where former teammates tend to work together again beyond what is expected from the perspective of work skills and supply and demand, aspects that are rightfully considered important in modeling work mobility in both open and internal labor markets~\cite{labor_flows,Guerrero-Frictional,occupations-automation,what-you-do,Moro2021}. However, the precise details of the reunion mechanism require further study. It is likely that a combination of effects take place to encourage this behavior. For example, individuals may be motivated to reunite because of a belief that this can improve team performance (e.g. in scientific research, more creative teams combine both newcomers as well as prior collaborators~\cite{Uzzi}). It is likely that referrals are an important enabling mechanism behind our observations~\cite{referrals-review}. Furthermore, it would be critical to assess in which ways reunions may enhance, hinder, or be neutral to team performance, a research direction that will require novel data.

Our work also introduces a new way to conceptualize the identity of a team that may aid in our understanding of real teams embedded in organizations. Concretely, because teams are so dynamic in their composition and because those dynamics can be driven both by individuals within the team as well as organizational decisions external to a team, we find it necessary to formulate a conceptual framework that provides a team with an identity as long as team member turnover is not driven by external influence. These ideas suggest the introduction of the notions of coordinated and uncoordinated moves into and out of teams. Although our concrete solution is particular to our setting and the data we work with, our notions are not isolated and, in fact, relate to classic ideas in the study of small groups~\cite{mcgrath_teams}. We note that our emphasis here is to study teams undergoing uncoordinated moves in order to develop a better picture of the steady employee-driven job mobility in an entire organization, but a broader look at this problem that includes coordinated moves and teams that change identities is necessary and will be part of future research.  

We believe the connection between teams and reuniting moves is fundamental. This is grounded in the fact that it is unlikely that people would form a close personal bond with others in settings where too many people work. Hence, although the definition of reuniting moves could be applied to a large division of a large organization, it is likely that most people within such large groups do not know each other or, if they do, that they work in any close manner. Furthermore, if reuniting moves are counted within a large grouping of personnel, the results can be misleading (e.g. in the limit were an entire organization is considered one team, all internal moves would count as reuniting moves). Thus, we posit that reuniting moves are fundamentally appropriate to small team contexts and are driven by collaboration and, probably, referrals.  

At the broad level of organizational careers, we believe our formulation of the problem of organizational job change is not only novel, but also capable of opening new lines of study in the research areas of organizational labor markets~\cite{white_1970,stewman} and manpower analysis~\cite{manpower} that have so far limited their approaches to less structured understanding of organizations. In our framework, many features that have been missing are added: teams, temporal evolution, occupational specialization, and social interactions are all represented and thus provide a large-scale granular picture of the highly complex organizational system. In addition, given the continued increased of data sources about organizations, driving the growth of practices and research such as personnel (or HR) analytics~\cite{Marler02012017}, it is our expectation that our approach would be applicable by others. Even the possibility of organizational modeling (see Ref.~\cite{carley2014computational}) is open under reasonable assumptions such as mimicking the structure of the organization of interest and assuming social interaction and other details as dictated by the situation. In essence, we expect these finding to offer a guide for a new generation of precise mobility models that include mechanisms long neglected that, as we see here, play crucial roles.

Although we have constructed our results by analyzing the AAW, the features of the teams we find through our method are consistent with those found in the rest of the literature on teams. For instance, we note that in the literature on teams many empirical meta-analyses observe teams of sizes ranging from two (2) to (10) members very often. For example, in Ref.~\cite{Horwitz} out of 27 studies they reviewed, no organization had teams with more than ten members. Larger team sizes can be found in some settings~\cite{hirschfeld2008mental} but are not common. With regards to the temporal variable, meta-analyses have found team lifetimes to typically be between one to two years~\cite{Bell}, and it is known that team members cycle through teams even faster than the existence of the team~\cite{Mathieu}. This also matches our findings. Combined with the fact that the social mechanism is a well established one across the literature, both in the context of job search~\cite{strength_of_ties} and organizational mobility~\cite{podolny1997resources}, we believe that our results are likely to hold broadly even if the specific percentages of reuniting moves can be somewhat different. However, we do acknowledge that our framework should not be taken to be universally applicable; some employers with different structures of job mobility (e.g. universities), may not conform to this framework entirely.

One additional question we must contend with is the disruption that organizations may suffer from new technologies. In this context, our expectation is that organizations will continue to have sizes relatively similar to those they have today, with the caveat that some workplaces may diminish in size and others may potentially increase~\cite{Alexopoulos,Gali,shea1998technology,Rio-Chanona}. However, the emergence of large language models (LLMs) are likely to have multiple effects that include (i) the replacement of some human tasks with LLM tools, (ii) the emergence of teams constituted by combinations of people and LLMs, and (iii) the implementation of organizational strategic decision-making aided by LLMs and, more broadly, artificial intelligence tools. Over some time horizon, these changes will find their productivity sweet-spot for their organizations, but we believe that the phenomenon of social bonding we observe in the current study will likely be disrupted only to a level that enters into equilibrium with the effects of the new technologies. At some point, in teams that require combinations of human-to-human and human-machine interactions~\cite{tsvetkova2024newsociology}, if new technologies disrupt them too much, it is likely that this will diminish optimal organizational output by way of disturbing the productive value of the social bonds within the teams. Thus, we believe that new technologies will reconfigure but not eliminate how social bonds will operate in the future, including the tendency for people to seek to reunite with prior coworkers.

In summary, in this article we introduce a novel framework of organizational modeling, centered around teams, that illuminates the effects of social bonds in organizational job moves. This social effect alone opens the possibility of designing models of organizations and their personnel that might eventually lead to rich forms of organizational career forecasting. Our approach is compatible with evolving organizations, an aspect of organizational job mobility modeling that has received little attention.

\bibliography{refs}
\bibliographystyle{unsrt}

\appendix
\subsection*{Availability of Data}
Due to the proprietary nature of the data, these cannot be shared without the express authorization from the US Department of the Army. We do provide a sample of data with limited fields to allow the replication of creating teams within an organization. The data does not have the same statistical properties of the real data as this would constitute an unathorized data release. The accompanying code for creating teams is provided as well. Data and relevant code for this research work are stored in GitHub:  \url{https://github.com/bryaneadams/Organizations-teams-and-job-mobility} and have been archived within the Zenodo repository:  \url{https://zenodo.org/records/16794752}. The release of the study has been approved by the US Army Acquisition Support Center (USAASC).

\subsection*{Competing interests}
The funders of the work (Army Research Institute) have played no role in the the design of the study; in the collection, analyses, or interpretation of data; in the writing of the manuscript; or in the decision to publish the results. The Office of the Director of Acquisition Career Management of the US Army has collaborated in this research but has not influenced the analyses or interpretation of the data. The study design is such that no elements of the data and analyses could compromise personal privacy or US national security.

\subsection*{Funding}
This work has been funded by the Army Research Office under contract W911NF2020117 and the Army Research Institute under contract W911NF2210250.

\subsection*{Author's contributions}
Conceptualization, B.A., V.V., E.L.; methodology, B.A., V.V., E.L., D.S. and M.P.; software, B.A., V.V., E.L., M.P.; data analysis, B.A., V.V., E.L.; writing—original draft preparation, B.A., V.V., D.S., E.L.; supervision, E.L. All authors have read and agreed to the published version of the manuscript.

\subsection*{Acknowledgements}

We acknowledge helpful discussions with Kyle Emich and Omar Guerrero. We also thank the Office of Research Computing (ORC) at George Mason University for use of its computational resources. 

\newpage

\appendix
\begin{center}
\textbf{\LARGE \sc Supplementary Information}
\end{center}

\renewcommand{\thetable}{S\arabic{table}}
\renewcommand{\thefigure}{S\arabic{figure}}
\renewcommand{\thesection}{S\arabic{section}}

\setcounter{figure}{0}
\setcounter{table}{0}

\section{Team characteristics}\label{sec:SI-team-stats}
\begin{figure}[t]
    \centering
        \includegraphics[width=0.8\textwidth]{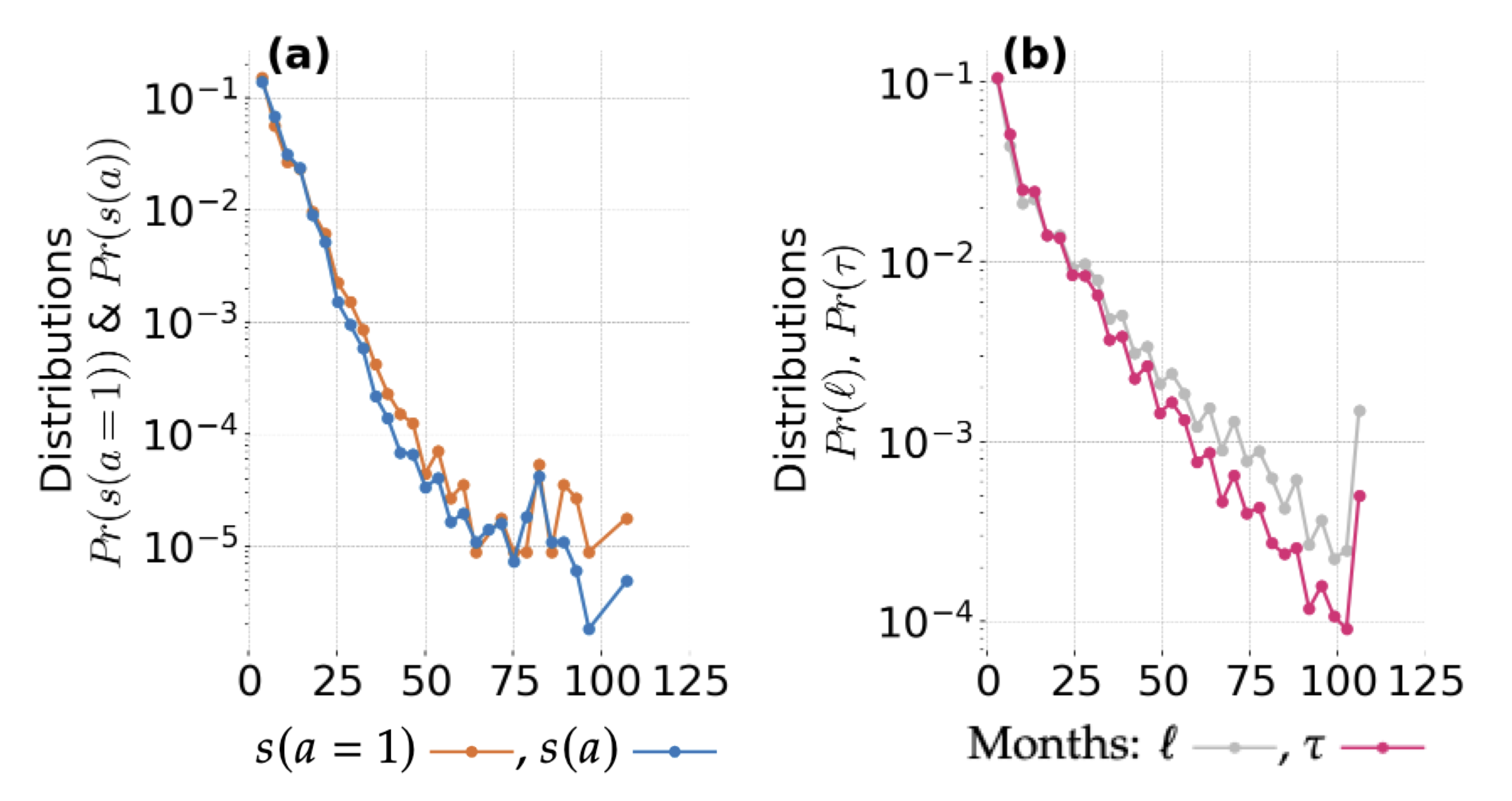}
    \caption{Statistical properties of continuing teams in the AAW.
    Panel \textbf{(a)} shows the size $a(a=1)$ distribution of teams (\protect\tikz[baseline=-0.5ex]{
    \protect\draw[wu_color, thick] (0,0) -- (.5,0);
    \protect\fill[wu_color] (0.25,0) circle (2pt);
}) at their initial age $a=1$, which decays exponentially until approximately team sizes of 50 or greater. The panel also shows the size distribution for $s(a)$, i.e. all sizes held by a time over their respective entire lifetime $\ell$ (\protect\tikz[baseline=-0.5ex]{
    \protect\draw[uic_color, thick] (0,0) -- (.5,0);
    \protect\fill[uic_color] (0.25,0) circle (2pt);
}). From this size, the distribution crosses over to a more even shape partly reflecting the small number of teams of such sizes. Panel \textbf{(b)} shows the distribution of team lifetimes $\ell$ (\protect\tikz[baseline=-0.5ex]{
    \protect\draw[ocs_color, thick] (0,0) -- (.5,0);
    \protect\fill[ocs_color] (0.25,0) circle (2pt);
}) and the time $\tau$ a person spends in a team (\protect\tikz[baseline=-0.5ex]{
    \protect\draw[wul_color, thick] (0,0) -- (.5,0);
    \protect\fill[wul_color] (0.25,0) circle (2pt);
}). The distributions are very similar, showing that most people remain in their teams for as long as the latter exist.}
    \label{fig:wu_statistics}
\end{figure}

The statistics of AAW organizational teams and their members can be [studied from the results displayed in Fig.~\ref{fig:wu_statistics}. In the figure, we show the distributions of team sizes $s$ measured in two different ways (Fig.~\ref{fig:wu_statistics}(a)), team lifetimes $\ell$ (Fig.~\ref{fig:wu_statistics}(b)), and the time of service of members in their respective teams (team member-tenure $\tau$, also in Fig.~\ref{fig:wu_statistics}(b)). These distributions include all unique teams identified through the $9$-year range of our data, and encompass a total of 31,697 unique teams. 

\begin{figure}[t]
    \centering
    \includegraphics[width=0.75\textwidth]{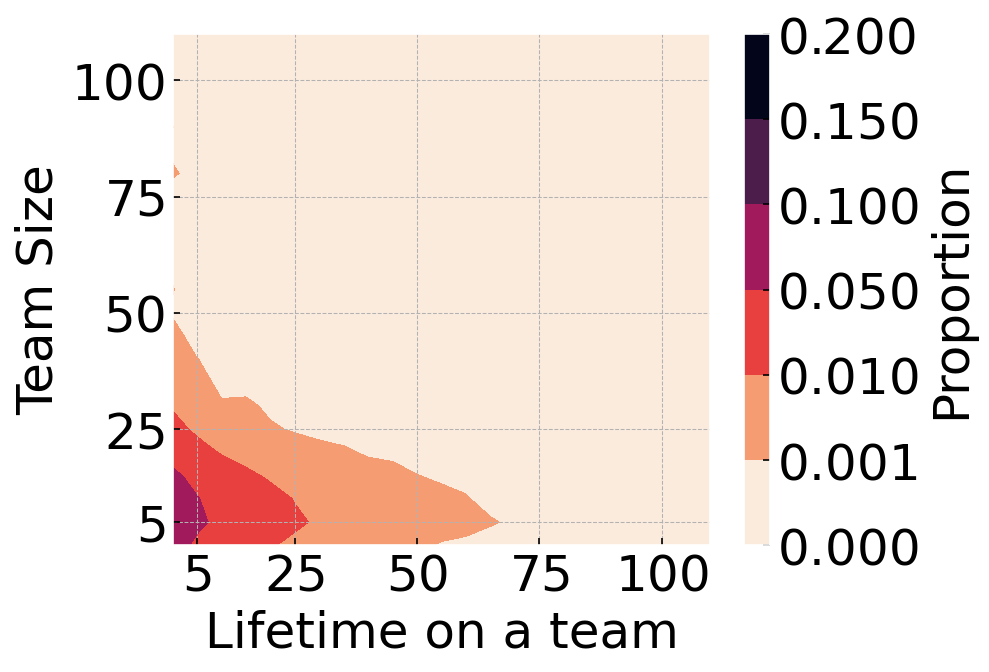}
    \caption{The heat map represents the density distribution of a person's lifetime on a team given a team size. The smaller the team, the longer a person remains on the team. This distribution demonstrates that large teams appear to be transitional entities, their lifetimes being considerably biased to the small limit ($\ell\lessapprox 5\ \text{months}$).}
    \label{fig:time_on_team}
\end{figure}

The team size distribution is measured in two ways. First, each unique team has a size $s$ that may change throughout the team's existence. We introduce the team age $a$, bounded from above by the team's overall lifetime $\ell$. At age $a$, the team has size $s(a)$. This generates $\ell$ distinct values of $s$ for the same team, and its distribution for all teams can be seen in Fig.~\ref{fig:wu_statistics}(a). However, team size typically has little variation over time and therefore, using the initial team size $s(a=1)$ is generally just as informative; this is the second distribution we present in Fig.~\ref{fig:wu_statistics}(a). Both distributions decay exponentially from $s=2$ to $s\approx 50$ members (where the distribution ceases to be exponential) as evidenced by the linear ranges of the curves in linear-log scale for increasing values of $s(a)$ and $s(a=1)$. The probability of teams with $s\gtrapprox 50$ is below $1$ in $10,000$. Additional analysis (Fig.~\ref{fig:time_on_team}) shows that such large teams appear to be transitional entities, their lifetimes being considerably biased to the lower limit ($\ell\lessapprox 5\ \text{months}$). However, these are so rare over the organization that their impact in our results is negligible. The mean team size is $\langle s(a)\rangle\approx 7.16$ people (which includes each team's supervisor).

In Fig.~\ref{fig:wu_statistics}(b), we focus on the temporal dimension. The team lifetime distribution is shown in orange and has a mean value of $\langle\ell\rangle\approx 14.59$ months. The employee team-tenure distribution appears in blue, and has an average of $\langle\tau\rangle\approx 12.14$ months, slightly smaller than team lifetime as would be expected, and providing evidence that most people remain in their teams for the entirety of the team's existence.

The behavior of very large teams ($s\gtrapprox 50$) can be understood from its relation to lifetime. In Fig.~\ref{fig:time_on_team}, we show a heat map version of the joint distribution of team size and and the tenure of its members, a reliable proxy for team lifetime as can be appreciated from Fig.~\ref{fig:wu_statistics}(b). From it, we observe that generally smaller teams have the largest tenures and that, as team size increases, tenure decays. 


\section{Time window effects}
The parameter $\Delta t$ before and after a time point $t$ determines the time window within which we test consistency of job moves into and out of a team. In the main text, we have chosen $\Delta t=6$ months as appropriate for two reasons. First, for the AAW, $2\Delta t \approx \langle \tau\rangle$, i.e. the parameter $\Delta t$ is such that the time window $2\Delta t$ matches the time a team member belongs to a team. Second, which may not be unrelated, in broader employment contexts yearly periodicity in the hazard rate of job separation has been observed~\cite{Farber}. However, it is still informative to explore other values of $\Delta t$ to confirm the robustness of our results. 

A change in $\Delta t$ directly affects the numbers and proportions over total of several quantities, such as (i) continuing teams, (ii) the people in continuing teams, and (iii) person-months in continuing teams. Indirectly, these changes may affect the quantities we have studied in this work. In the next sections, we explore these direct and indirect effects. 

\subsection{Overall AAW team personnel coverage by $\Delta t$ selection}
Few teams span the duration of our data (see Fig.~\ref{fig:wu_statistics}(b)) and an increase in $\Delta t$ should lead to a decrease in the number of continuing teams. We first assess how much change there is as we modify $\Delta t$ in terms of the numbers of continuing and non-continuing teams (teams that are born or die within the time window), the number of workers these types of teams have, and the number of person-months associated with types of teams. We study three values of $\Delta t$, 6, 7, and 9 months, corresponding to time windows of $2\Delta t=12$, $14$, and $18$ months. The first value is chosen to match $\langle\tau\rangle$, the second matches $\langle\ell\rangle$, and the last value is chosen to provide perspective for time windows that exceed both member tenure and team lifetime. 

\begin{figure}[t]
    \centering
    \includegraphics[width=0.8\textwidth]{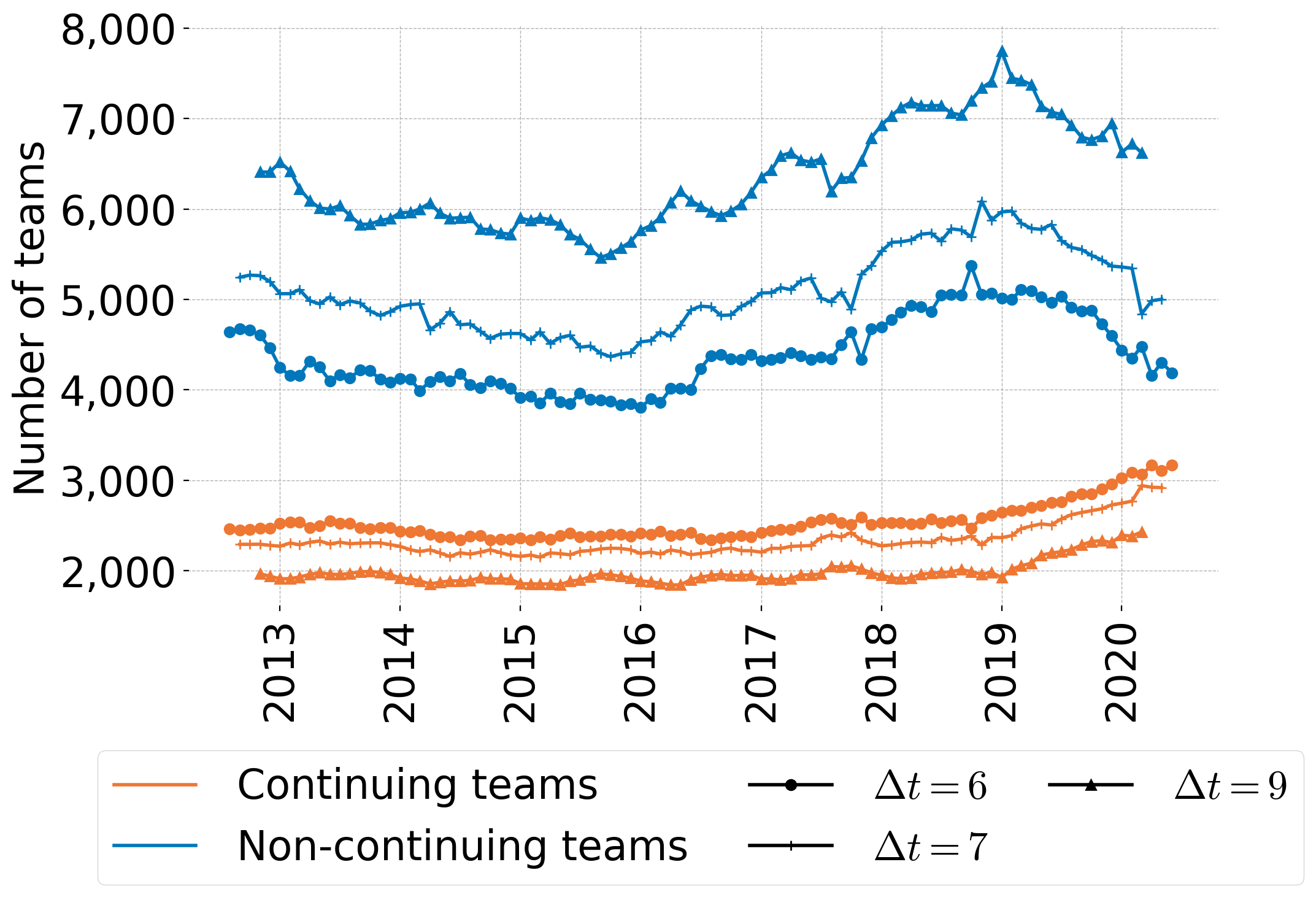}
    \caption{Time series for the numbers of continuing  (\protect\tikz[baseline=-0.5ex]{
    \protect\draw[wu_color, ultra thick] (0,0) -- (.5,0);
}) and non-continuing (\protect\tikz[baseline=-0.5ex]{
    \protect\draw[uic_color, ultra thick] (0,0) -- (.5,0);
}) teams for different values of $\Delta t$. As $\Delta t$ increases, the curves providing the numbers of continuing teams move down at a slower rate than the curves for non-continuing teams move up.}
    \label{fig:team_coverage}
\end{figure}
In Fig.~\ref{fig:team_coverage}, we present the number of continuing and non-continuing teams as a function of $t$ for the three values of $\Delta t$. As expected, the curves providing the numbers of continuing teams move down as $\Delta t$ increases, while the curves for non-continuing teams move up. It is useful to note that the decrease in continuing teams is slower than the increase for non-continuing teams. 

\begin{figure}[t]
    \centering
    \includegraphics[width=0.8\textwidth]{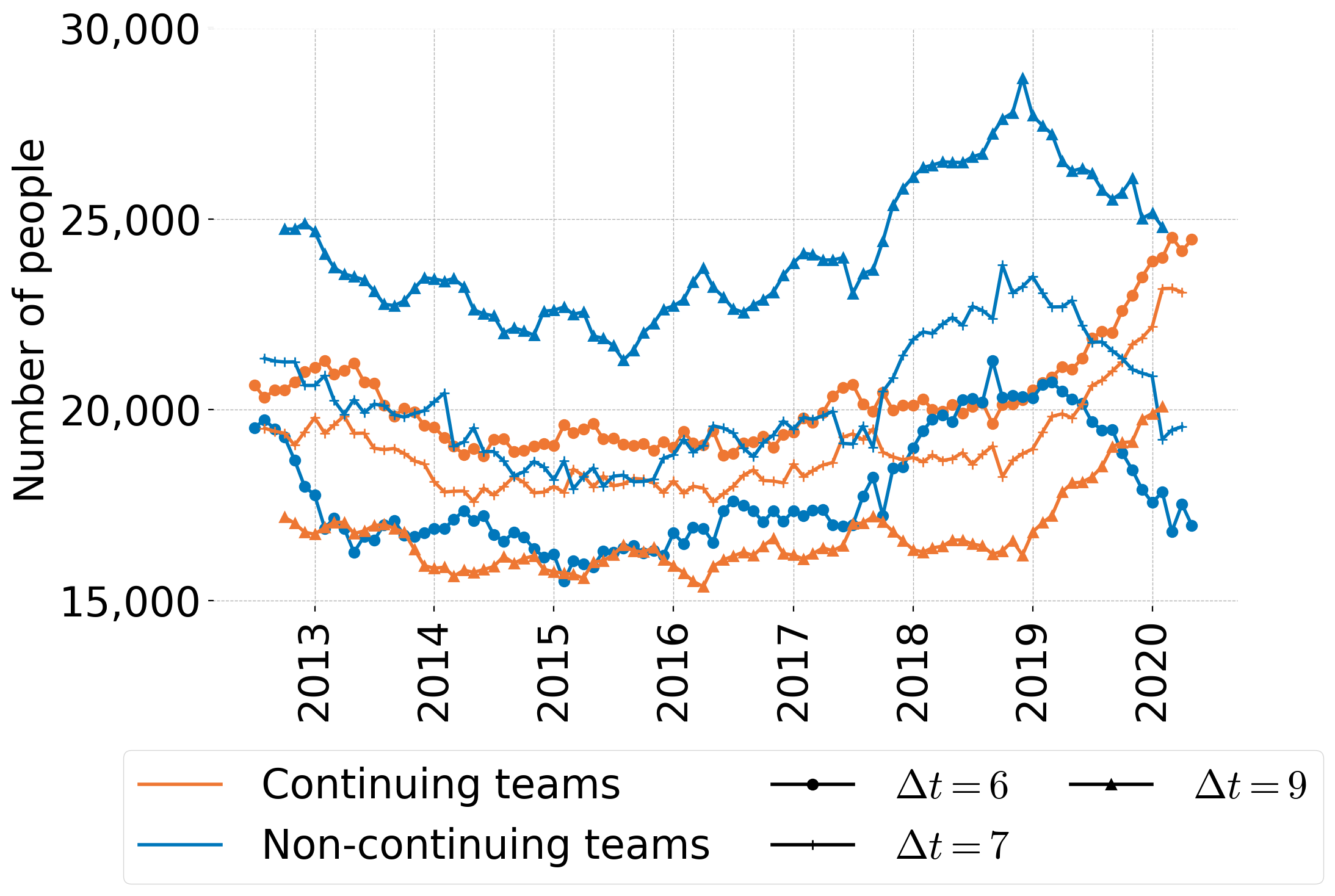}
    \caption{Time series of the number of employees associated with continuing (\protect\tikz[baseline=-0.5ex]{
    \protect\draw[wu_color, ultra thick] (0,0) -- (.5,0);
}) and non-continuing (\protect\tikz[baseline=-0.5ex]{
   \protect\draw[uic_color, ultra thick] (0,0) -- (.5,0);
}) teams for different values of $\Delta t$. As $\Delta t$ increases, the curves representing the number of employees associated with continuing teams decrease in value while those associated with non-continuing teams increase. This follows the same trends as seen in Fig.~\ref{fig:team_coverage}.}
    \label{fig:num_people}
\end{figure}
The number of employees associated with continuing and non-continuing teams follow the same trends as the numbers of teams (Fig.~\ref{fig:num_people}). Namely, the number of employees belonging to continuing teams decreases as $\Delta t$ increases, and vice versa for employees that are part of non-continuing teams. These results are consistent with the fact that team sizes are within a narrow distribution, as seen in Fig.~\ref{fig:wu_statistics}. It is interesting to note that when $\Delta t=7$ months, the numbers of team members belonging to the two different types of teams become roughly comparable. 

\begin{figure}[t]
    \centering
    \includegraphics[width=0.8\textwidth]{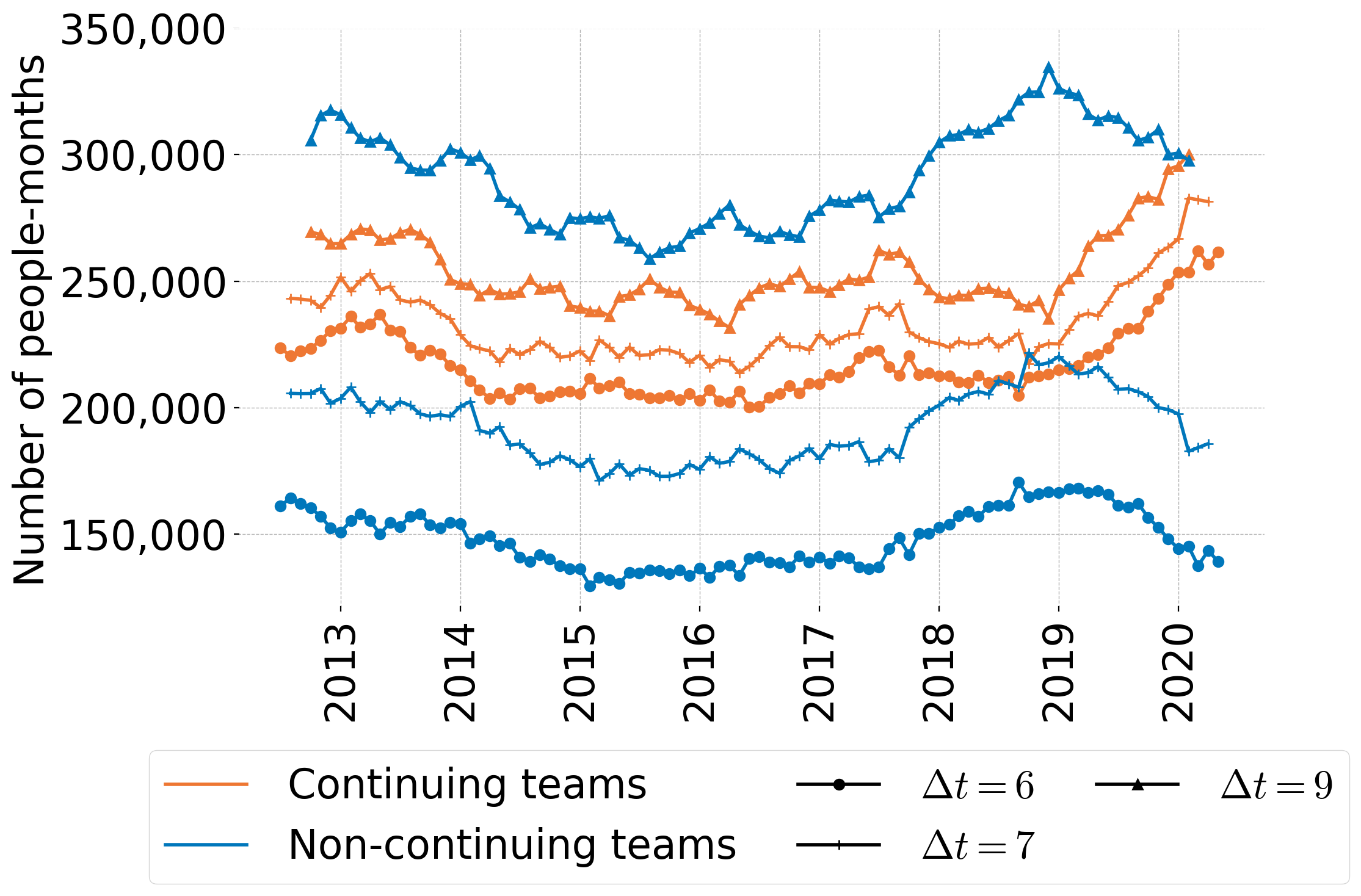}
        \caption{Time series of the numbers of people-months for employees in continuing (\protect\tikz[baseline=-0.5ex]{
    \protect\draw[wu_color, ultra thick] (0,0) -- (.5,0);
}) and non-continuing (\protect\tikz[baseline=-0.5ex]{
   \protect\draw[uic_color, ultra thick] (0,0) -- (.5,0);
}) teams for different values of $\Delta t$. In contrast to the trends found in Figs.~\ref{fig:team_coverage} and~\ref{fig:num_people}, an increase in $\Delta t$ also leads to an increase in person-months. This stems from the fact that the months a person spends in a team are monotonic with $\Delta t$. There exists a sharp increase of non-continuing team person-months between $\Delta t =7$ and $9$ months. This is associated with $\langle \ell \rangle \approx 14$, which means that continuing teams with $2\Delta t = 18$ months are rare.}
    \label{fig:people_months}
\end{figure}
One last check of the direct effect of $\Delta t$ is the number of person-months of people working in continuing or non-continuing teams. The analysis is exhibited in Fig.~\ref{fig:people_months}. The behavior of these curves is different than that of the two previous analyses because the time increase contributes to an increase in the months that make up the person-months. Consequently, both continuing and non-continuing person-months curves increase along with $\Delta t$. However, the interesting observation is the sharp increase of non-continuing team person-months between $\Delta t=7$ and $9$ months. This is associated with $\langle\ell\rangle\approx 14$ months, which means that continuing teams with $2\Delta t=18$ months are atypical and therefore more or less uncommon. In other words, for a time window of $2\Delta t=18$ months, employees mostly deliver their work effort into teams that have changed their identity at some point within that window. In contrast, for $2\Delta t=14$ months, that effort is still mostly delivered within continuing teams.

\subsection{Robustness checks}
We now study the effect of $\Delta t$ on $z_{m,n}$, $y_{m,n}$, and $\lambda_{m,n}$. 

\begin{figure}
    \centering
    \includegraphics[width=0.8\linewidth]{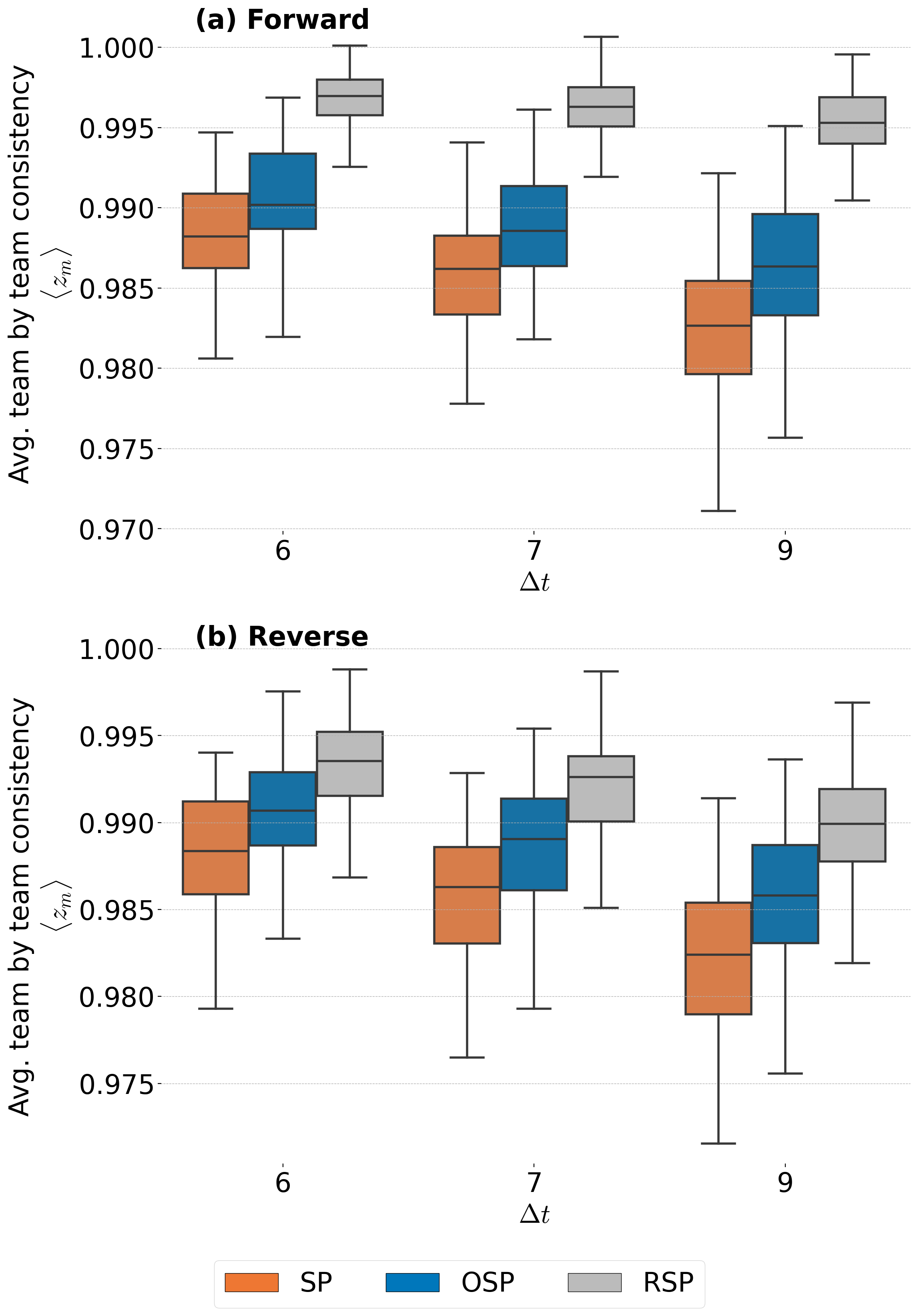}
    \caption{Box plots of team by team forward (panel \textbf{(a)}) and reverse (panel \textbf{(b)}) average consistency $\langle z_{m}\rangle$. The box plots are constructed under the same rules as those in Fig.~\ref{fig:zm} but organized from left to right with $\Delta t=6$, $7$, and $9$ months.}
    \label{fig:zm_reverse_robust}
\end{figure}
First, we revisit forward and reverse team by team consistency $z_{m,n}$ for the three models (Fig.~\ref{fig:zm_reverse_robust}). The qualitative features of our analysis continue to hold which are that for all three values of $\Delta t$, the models continue to perform in the same order, with RSP being the best performing, OSP the second best, and SP the worst. In addition, we note a very slight decrease in consistency as $\Delta t$ increases, of the order of $1\%$, although consistency of the RSP model is the least affected. Since the trend is very small and the values of consistency are already close to 1, we do not explore its origin. 

\begin{figure}
    \centering
    \includegraphics[width=0.8\linewidth]{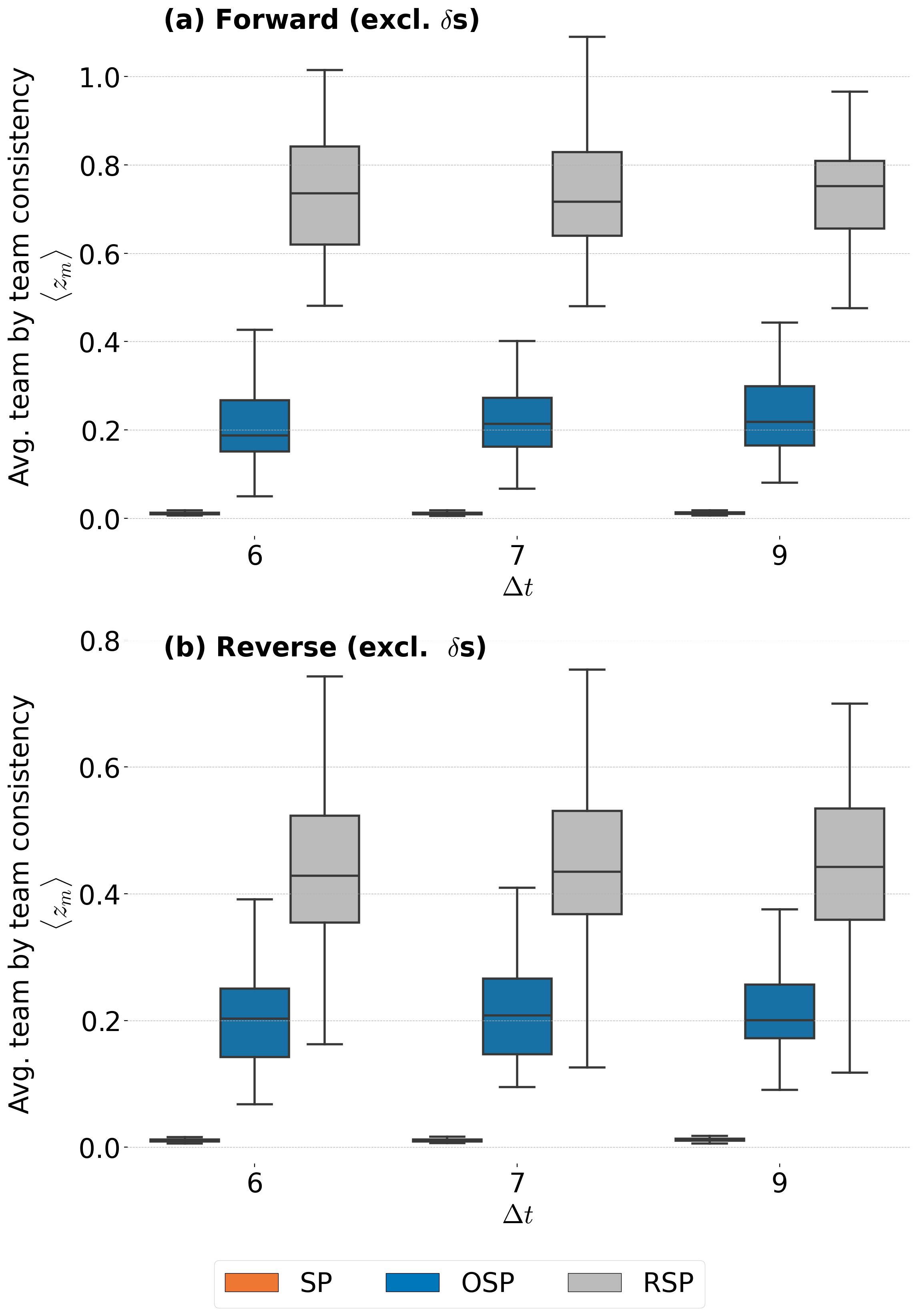}
    \caption{Box plots for team by team forward (panel \textbf{(a)}) and reverse (panel \textbf{(b)}) average consistency $\langle z_{m}\rangle$ with all continuing teams remove that are without transitions during the entire time window $t-\Delta t$ to $t+\Delta t$. The box plots are constructed under the same rules as those in Fig.~\ref{fig:zm} but organized from left to right with $\Delta t=6$, $7$, and $9$ months.}
    \label{fig:zm_reverse_non_inflated_robust}
\end{figure}
Similar behavior is found for the forward and reverse team by team consistency $z_{m,n}$ for the three models that exclude the teams without labor moves, i.e. for the use of Eq.~\ref{eq:z_m} while ignoring the Kronecker delta (Fig.~\ref{fig:zm_reverse_non_inflated_robust}). In this case, there is hardly any discernible difference between models or time periods. This represents very strong evidence to the robustness of the influence of reuniting moves in job mobility.

\begin{figure}
    \centering
    \includegraphics[width=0.8\linewidth]{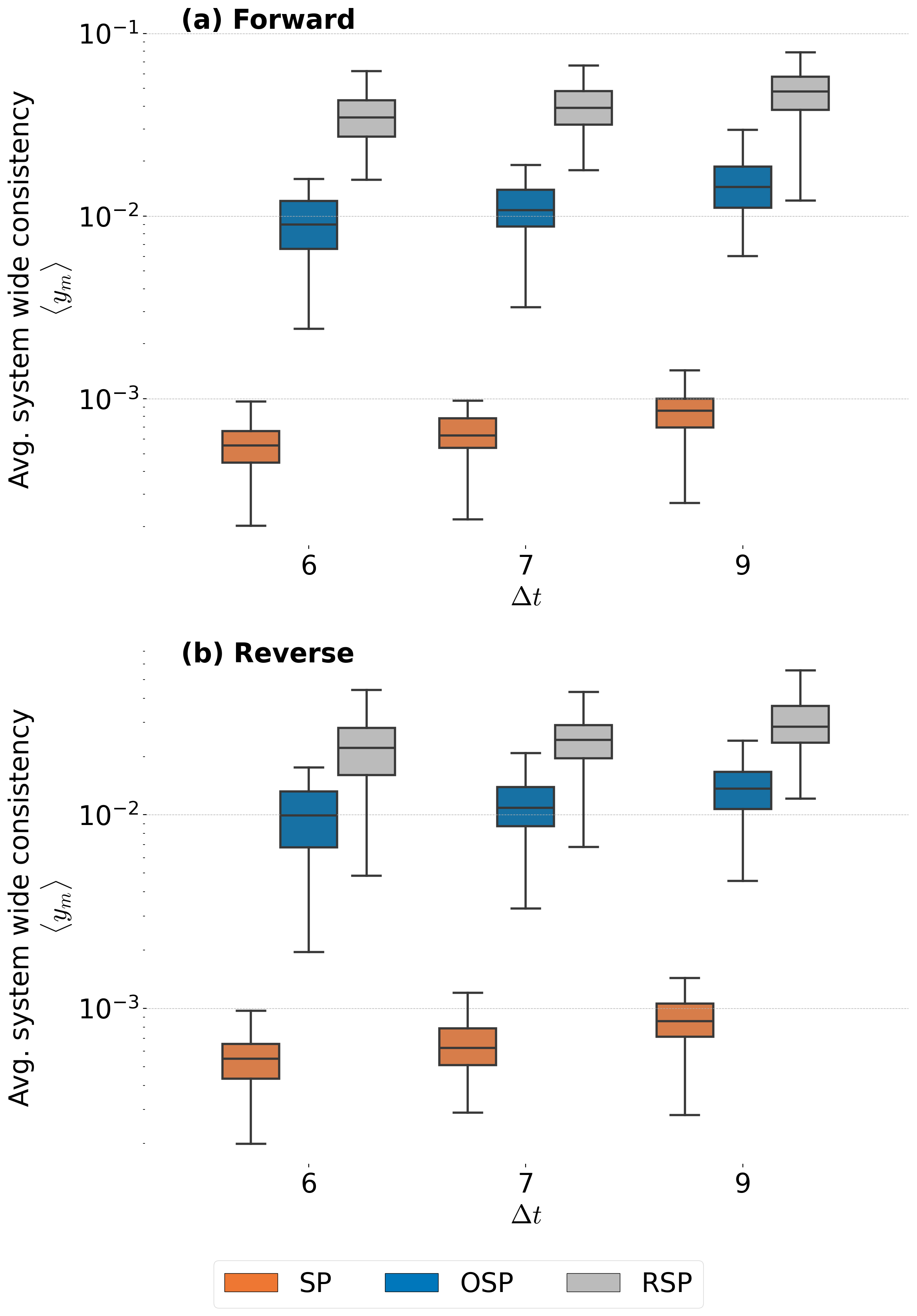}
    \caption{Box plots for system-wide forward (panel \textbf{(a)}) and reverse (panel \textbf{(b)}) average consistency $\langle y_{m}\rangle$. The box plots are constructed under the same rules as those in Fig.~\ref{fig:ym} but organized from left to right with $\Delta t=6$, $7$, and $9$ months.}
    \label{fig:ym_reverse_robust}
\end{figure}
Qualitatively, the behavior of system-wide forward and reverse consistency $y_{m,n}$ is virtually the same across $\Delta t$ (Fig.~\ref{fig:ym_reverse_robust}). There is a slight increase in performance with increasing $\Delta t$ which again supports the importance of reuniting moves in narrowing down the possible job changes that the organization can experience.

\begin{figure}
    \centering
    \includegraphics[width=0.8\linewidth]{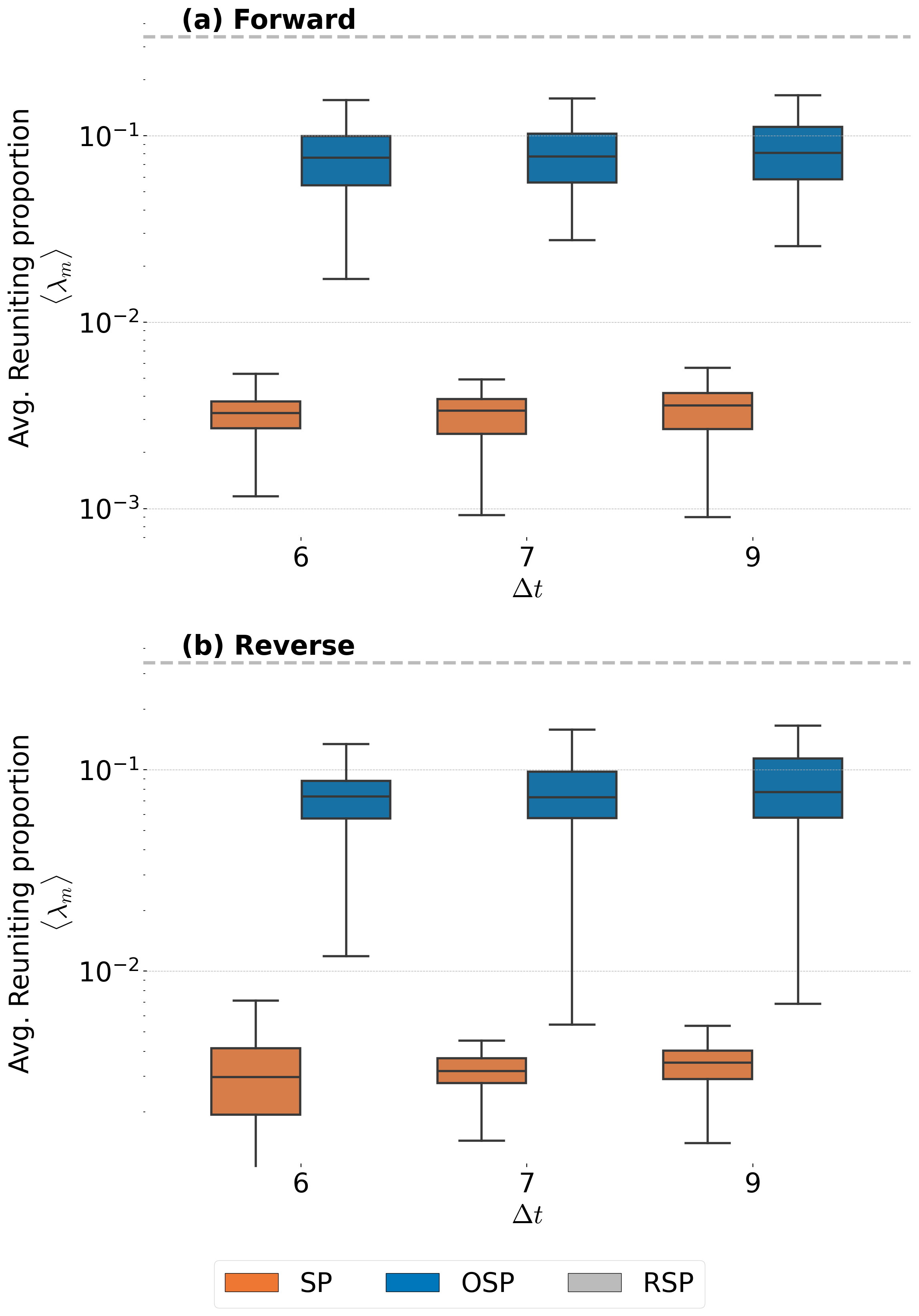}
    \caption{Box plots for the forward (panel \textbf{(a)}) and reverse (panel \textbf{(b)}) ratios of observed versus random reuniting moves from the $\text{SP}$ and $\text{OSP}$ models, $\langle\lambda_m\rangle$. The box plots are constructed under the same rules as those in Fig.~\ref{fig:per_follow} but organized from left to right with $\Delta t=6$, $7$, and $9$ months.}
    \label{fig:l_m_reverse_robust}
\end{figure}
Finally, the proportion of reuniting moves by models OSP and SP does not significantly change as $\Delta t$ changes (Fig.~\ref{fig:l_m_reverse_robust}), which indicates that varying the time window is not able to make these two models better predict the reuniting patterns observed in the real system. 

From these analyses, the main conclusion we are able to draw is that reuniting job moves are a robust feature of internal job mobility and which carry more information about the system than job moves that preserve occupational mobility or labor supply and demand. Furthermore, our exploration of the parameter $\Delta t$ does not indicate that our conclusion is an artifact of the way in which we select continuing teams on the basis of their longevity.

\clearpage

\section{Job transition demographics}

\begin{figure}[H]
    \centering
    \includegraphics[width=0.8\textwidth]{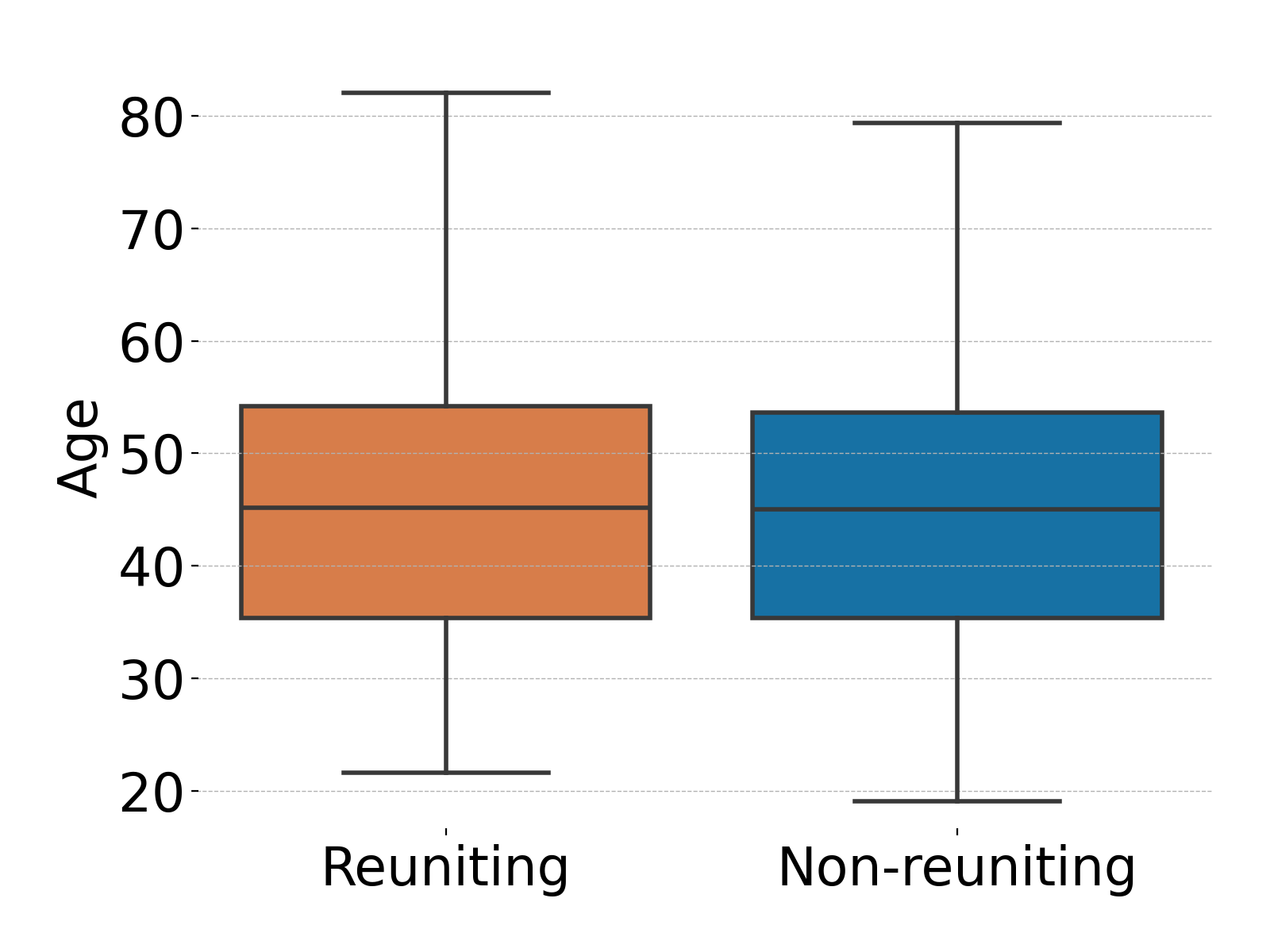}
    \caption{Box plots for the age of AAW employees performing reuniting versus non-reuniting uncoordinated moves. The average age of employees performing a reuniting move is 45.0 years, while non-reuniting moves is 44.7 years. The closeness of these values signals age does not influence a person's likeliness of conducting a reuniting move.}
    \label{fig:age_of_movers}
\end{figure}

\begin{figure}[H]
    \centering
    \includegraphics[width=0.8\textwidth]{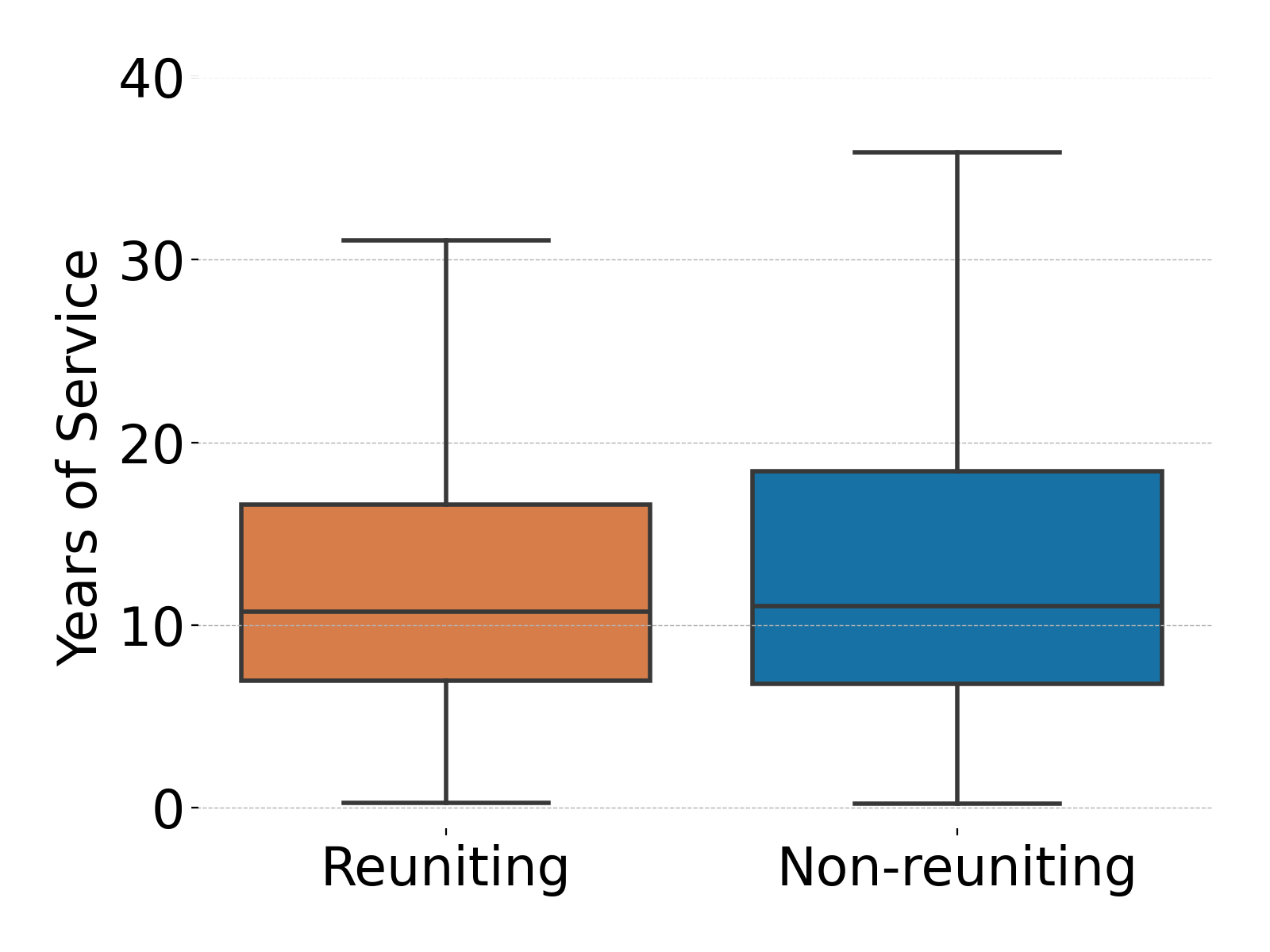}
    \caption{Box plots for the years of government service of AAW employees performing reuniting versus non-reuniting uncoordinated moves. The average years of government service of employees performing a reuniting move is 13.2 years, while non-reuniting moves is 13.7 years. The closeness of these values signals years of government service does not influence a person's likeliness of conducting a reuniting move.}
    \label{fig:yos_of_movers}
\end{figure}

\begin{figure}[H]
    \centering
    \includegraphics[width=0.8\textwidth]{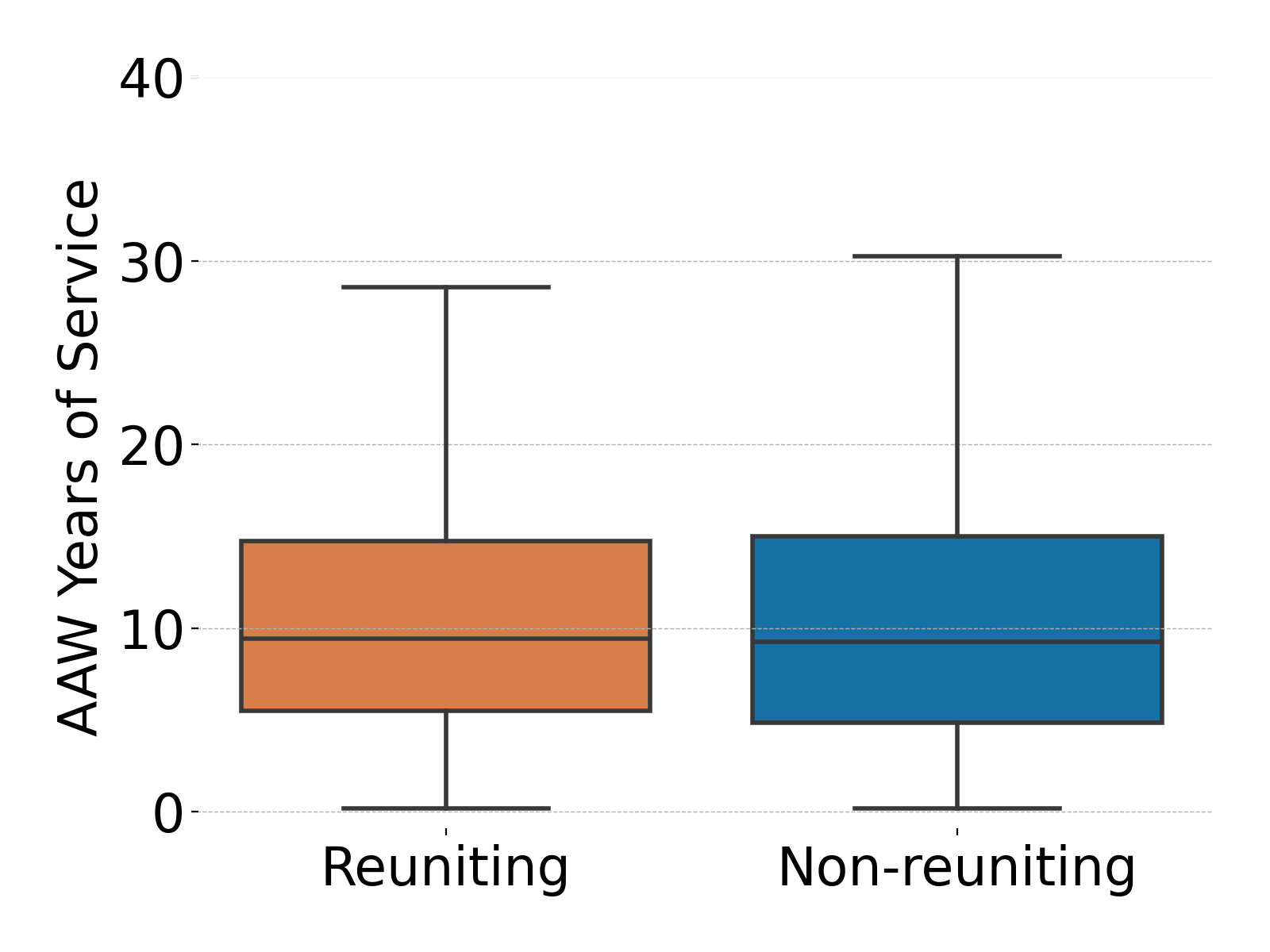}
    \caption{Box plots for the years of AAW service of AAW employees performing reuniting versus non-reuniting uncoordinated moves. The average years of AAW service of employees performing a reuniting move is 11.2 years, while non-reuniting moves is 11.0 years. The closeness of these values signals years of AAW service does not influence a person's likeliness of conducting a reuniting move.}
    \label{fig:aaw_yos_of_movers}
\end{figure}

We explored demographic variables (age, gender, ethnicity, years of service to the government, and AAW) for all uncoordinated moves within the AAW and did not find any significant relationship between these variables and the likelihood of performing a reuniting versus non-reuniting move within the AAW. Reuniting moves, both coordinated and uncoordinated, make up 32\% of all moves between teams, slightly less than that of uncoordinated reuniting moves between continuing teams (which occur at a rate of $34\%$).  Figs.~\ref{fig:age_of_movers},~\ref{fig:yos_of_movers}, and~\ref{fig:aaw_yos_of_movers} show that there are no tendencies for younger or more junior employees, as measured by years of service, to perform an uncoordinated reuniting move more often than more senior employees. Tabs.~\ref{tab:gender_movers} and~\ref{tab:ethnicity_movers} provide proportions of reuniting moves by gender and ethnicity, respectively (see details in the captions). Each category aligns with the proportion of uncoordinated moves found in the AAW, signaling no observed impact on the likelihood of conducting a reuniting transition.

\begin{table}[H]
    \centering
    \begin{tabular}{lcc}
        \toprule
        Gender & Reuniting & Non-reuniting \\
        \midrule
        Female & 3149 (30.6\%) & 7137 (69.4\%) \\
        Male & 5105 (30.1\%) & 11878 (69.9\%) \\
        \bottomrule
    \end{tabular}
    \caption{Both female and males conduct the same proportion of reuniting uncoordinated moves. Both gender proportions of reuniting moves are similar to the overall proportion of reuniting moves in the AAW.\label{tab:gender_movers}}
\end{table}

\begin{table}[H]
    \centering
    \begin{tabular}{lcc}
        \toprule
        Ethnicity & Reuniting & Non-reuniting \\
        \midrule
        White (Non-Hispanic) & 2,731 (27.6\%) & 7,177 (72.4\%) \\
        Other & 1,213 (29.3\%)& 2,933 (70.7\%) \\
        Missing & 4,310 (32.6\%) & 8,905 (67.4\%) \\
        \bottomrule
    \end{tabular}
    \caption{A large portion of our data do not provide the ethnicities of their employee's; however, as with gender, the proportions of each ethnicity that conduct reuniting uncoordinated moves are similar to the overall proportion of reuniting moves in the AAW. This includes individuals missing ethnicity data.\label{tab:ethnicity_movers}}
\end{table}

\end{document}